\documentclass[twocolumn]{article}
\usepackage{graphicx} 
\usepackage{cite}
\usepackage{color}
\definecolor{dl}{rgb}{0.9,0.9,0.9}
\definecolor{orange}{rgb}{0.8,0.4,0}
\definecolor{purple}{rgb}{0.7,0,0.7}
\definecolor{darkgreen}{rgb}{0,0.4,0}
\definecolor{grey}{rgb}{0.75,0.7,0.7}

\begin{document}

  \title{A simple model system for self-propelled particles passing a bottleneck}

    \author{{Mahdieh Mohammadi$^{1,2}$}, {Kirsten Harth}$^1$, {Dmitry Puzyrev$^1$}, {Tina Hanselka}$^1$,\\
             Torsten Trittel$^1$, and { Ralf Stannarius$^1$}  \\
        $^1$ Institute of Physics, Otto von Guericke University Magdeburg,\\ 39106 Magdeburg, Germany \\
        $^2$ Department of Physics, Institute for Advanced Studies in Basic Sciences (IASBS),\\ Zanjan, Iran}

\date{\today}

    \maketitle

    \begin{abstract}
We study the passage of active and passive granular particles through a bottleneck under gravitational bias.
The grains are pharmaceutical capsules with nearly spherocylindrical shapes on a vibrating table. The vibrating plate is slightly tilted in order to break the in-plane symmetry and to give particles a motivation to move in a preferential direction on the plate. The passage through a narrow gate with openings comparable to the grain length is studied using video imaging, and particle positions and velocities are extracted from the recorded frames. We compare the behaviour of asymmetrically filled, active capsules with symmetrically filled, passive ones.
    \end{abstract}

\section{Introduction}

The evacuation of a crowd of people from a room, egress of animals from a corral and  discharge of
granular material from a silo are, for different reasons, very
active, interrelated topics of current research. Seeking the similarities and
differences in these  processes promises a more universal insight
into their underlying mechanisms. On the one hand, granular systems
can surely not mimic the complex, individual behavior of
spontaneously decision-making people or animals, but may nevertheless reproduce
fundamental features of egress studies. On the other hand, conducting
experiments or field studies with humans, or with animals, is
restricted by ethical principles and laws, requiring tremendous efforts and costs. There are usually restrictions to safe
situations (no panic) with comparatively few participants, or one has to refer to data of coincidentally recorded events. The
outcome is influenced by psychological components which are hard to
determine quantitatively, and  it usually depends on  social, cultural and professional background, as well as the age of the test persons~\cite{Frey2010,Sieben2017,Ramirez2019,Adrian2020}. Computer simulations,
succeeding the works of Helbing et al.~\cite{Helbing2000,Helbing2001,Helbing2002}, often
capture only a fraction of the complexity, or may require
substantial computation effort. This motivates studies of
less complex, inanimate objects that reproduce some important features of these processes. The properties of such mechanical objects are easier to
control, and experiments can be readily handled and reproduced in a
laboratory. One such example are vibration-driven granular
particles, e.~g.~\cite{Nicolas2018,Kudrolli2008,Kudrolli2010,Volfson2004,Gandra2019,Tsai2005,Deseigne2012,Scholz2018,Barois2019,Soni2020}. The usefulness of such self-propelled rod-like objects to mimic the behavior of biological systems in vitro has been comprehensively discussed recently in a review article by B\"ar et al. \cite{Baer2020}. A good overview of recent studies of pedestrian dynamics can be found in Refs.~\cite{Haghani2020a,Haghani2020b}.
A more general review of the dynamics of systems of active particles was provided by Bechinger et al.~\cite{Bechinger2016}.
For a comparison of living and granular systems see, for example, Refs. \cite{Zuriguel2014,Zuriguel2017}.

We will first recollect some similarities between the passage of living objects through narrow outlets and the discharge of grains from a silo.  Emptying of containers through their orifices can follow two different
dynamic scenarios~\cite{To2001,To2005,Zuriguel2005,Thomas2013}: In
the first one, which occurs at sufficiently large outlet widths, the
outflow is practically continuous until the container is practically empty,
without considerable flow rate fluctuations. The second one is
characterized by transient or permanent blockages of the outlet,
 usually undesired. Such clogging or strong fluctuations of the flow rate
cause severe problems, e.g. in  construction, agricultural, or pharmaceutical industries. In crowds of humans or animals, an even more severe
consequence is a build-up of high pressure near the
outlet~\cite{Pastor2015,Zuriguel2020,Wu2018,Helbing2012}, which may lead to casualties and
unresolvable blockages of the door.

In both biological and granular systems, not only the prevention
of clogs, but also a minimization of the total time needed for emptying
the room or container (the egress time) and the avoidance of strong pressure
fluctuations are of substantial importance.
 For people or animals moving through constrictions, 
the actual {\em motivation} of the
individuals is expected to strongly affect the pressure near the
outlet, as qualitatively confirmed by Pastor et
al.~\cite{Pastor2015}. Similar observations were made in experiments
with animals such as, e.~g., sheep, mice or ants~\cite{Shiwakoti2011,Soria2012,Zuriguel2014,Parisi2015,Garcimartin2015,Lin2016,Wang2018,Gwynne2018,Haghani2020a,Haghani2020b}, and in numerical simulations
~\cite{Lin2015,Wang2019,Parisi2018}. Obviously, this intrinsic motivation to pass the outlet is hard
to quantify or even to control in experiments with living creatures. For
sufficiently wide exits, without congestions, the outflow rate (in persons per door width $d$ and time$t$) is commonly assumed to be independent of the exit width~\cite{Liao2017,Haghani2020a,Haghani2020b} (but see also \cite{Gwynne2009}).

Whereas in living systems, the motivation to pass an outlet is set by a certain stimulus of the individuals, studies of granular material discharging from hoppers and silos typically rely exclusively on gravity forcing the grains toward the exit.
As long as a certain fill height of the container is exceeded, the outflow rate is independent of the fill level and satisfactorily described
by the Beverloo equation~\cite{Franklin1955,Beverloo1961,Nedderman1982,Mankoc2007,Anand2008}. This applies in the free flow regime as well as during avalanches
between clogs~\cite{Thomas2013}. It predicts an outflow rate that is proportional to $d^{1.5}$ in a 2D system.

The pressure at the bottom of the silo saturates above a certain height related to the container diameter, because forces are transferred to the side walls (Janssen
effect)~\cite{Janssen1895,Sperl2006}. The actual pressure depends on the geometry of the container near the outlet~\cite{Walters1973}.
The outflow and clogging characteristics are affected by inter-particle friction, particle shape and
deformability. A prominent feature of living systems that distinguishes them from grains in a classical hopper experiment is their activity and their intrinsic ability to perform
deliberate
rearrangements of their positions and orientations in situations
perceived unfavorable.

Some features of biological systems can be simulated by using
(almost) frictionless, deformable constituents, such as emulsion
drops~\cite{Hong2017a,Hong2017}  or
viscoelastic hydrogel
spheres~\cite{Harth2020,Hong2017,Ashour2017,Stannarius2019,Stannarius2019a}.
Then, permanent clogging occurs only at substantially smaller opening widths compared to hard frictional grains.
Interestingly, slow particle rearrangements in the container can evoke intermittent flow with non-permanent
congestions~\cite{Harth2020}, and the
pressure near the outlet can vary substantially in time - a
similarity to animals and pedestrians~\cite{Zuriguel2020,Pastor2015}.

Here, we report experiments with
elongated grains with specially prepared mass distributions, which are positioned on a flat plate mounted on an electro-mechanical shaker. We drive these spherocylinders by vertical vibrations of the plate. The driving frequency and amplitude set the diffusion parameters of the spherocylinders. At the same time, we break the in-plane symmetry of the shaker dish by tilting the setup by a few degrees, defining a preferred direction of particle motion. Particles are initially placed in a restricted area with a single exit.
The 'motivation' to pass this bottleneck exit
is controlled by the tilt angle of the setup that sets an effective
gravitational force towards the exit.
Experiments are conducted with symmetric grains that exhibit purely diffusive dynamics, and asymmetrically prepared, self-propelled grains.
We compare the egress behavior of the active grains to the
equally shaped but purely diffusive ones.
The behavior of this purely mechanical model system is compared to features of colloidal systems and the biological systems described above.

Before the experiment is described in detail, it is necessary to add
a few remarks to differentiate our experiment from experiments with
vibrated hoppers. Those, at first glance, appear to have some
resemblance to the present setup: Our sample container, viewed from
above, looks quite similar to a quasi two-dimensional hopper.
Indeed, the particles essentially remain in a monolayer. The important
difference is that gravity does not act primarily in the hopper plane. In our
setup, this plane is almost horizontal, nearly perpendicular to the
gravitational force. There is still some small effective gravitation
in the direction of the orifice when the shaker is tilted, which is
needed for the experiment. However, the effective gravity is too
small to overcome friction, so that the particles do not slide
downward when the shaker is off.

Several studies of vibrated hoppers
\cite{Wassgren1999,Wassgren2002,Mankoc2009,Janda2009b,Lozano2012,Zuriguel2017,Guerrero2018,Guerrero2019,Takahashi1968,Suzuki1968,Lindemann2000,Chen2006,Kumar2020}
probed the stability and the
destruction of clogs formed at the orifice,
denoted as unjamming. Some other studies applied vibrations during
silo discharge in order to characterize either the inhibition or the
support of discharge rates
\cite{Takahashi1968,Suzuki1968,Lindemann2000,Chen2006,Kumar2020}.
Regarding the present study, the consequences of the vibrations are considerable different. In all of these earlier
experiments, the grains are driven downward towards the orifice by
the gravitational force, and the vibrations in the same direction
break arches and prevent clogging. In our experiments, the
oscillations are perpendicular to the slope of the container and
serve only as the energy supply to generate diffusive or directed
particle motion.

Similar to typical
biological situations, the particles we use are substantially elongated. The prolate shape of the grains strongly affects their
diffusion character\-istics~\cite{Kudrolli2010,Yadav2012,Baskaran2008,Roh2015}. 
Shape-anisotropy can also affect the packing in denser ensembles (see e.g.
Refs.~\cite{Borzsonyi2013,Borzsonyi2012a,Borzsonyi2012,Ashour2017a,Borzsonyi2016,Szabo2018,Daniels2009}).
In particular, elongated constituents tend to prefer alignment towards the
exit~\cite{Borzsonyi2016,Szabo2018,Ashour2017a}.
Elongated shapes can suppress
motility-induced phase separation~\cite{vanDamme2019}. Caging effects cause a decrease of perpendicular diffusion compared to axial diffusion~\cite{Kudrolli2010,Daniels2009}. 

This manuscript is organized as follows: We first describe the experimental setup, the
particles and the image evaluation methods. Then, we characterize
the dynamics of individual active or passive particles on a
horizontal plate, in dependence of the excitation parameters. After
this, the outflow characteristics of the grains are characterized
for specific driving parameters and tilt. Representative
orientational properties near the orifice are shown. Finally, we
discuss our results in the context of typical exit scenarios from
the literature.

\section{Experimental setup, materials and methods}
\subsection{Particle preparation}

We employed two types of particles with identical shape and size: 'active' and 'passive' capsules. The particles are excited with a vibrating table, and their dynamic behaviour differs qualitatively in that the 'passive' particles show an anisotropic diffusive behaviour with preferential motion along their long axes, but without preferred sense of direction, while the 'active' particles exhibit an additional directed motion along their long axes. Details are discussed in Sec.~\ref{Sec:Single}.

\begin{figure}[htbp]

        \includegraphics[width=0.4\columnwidth]{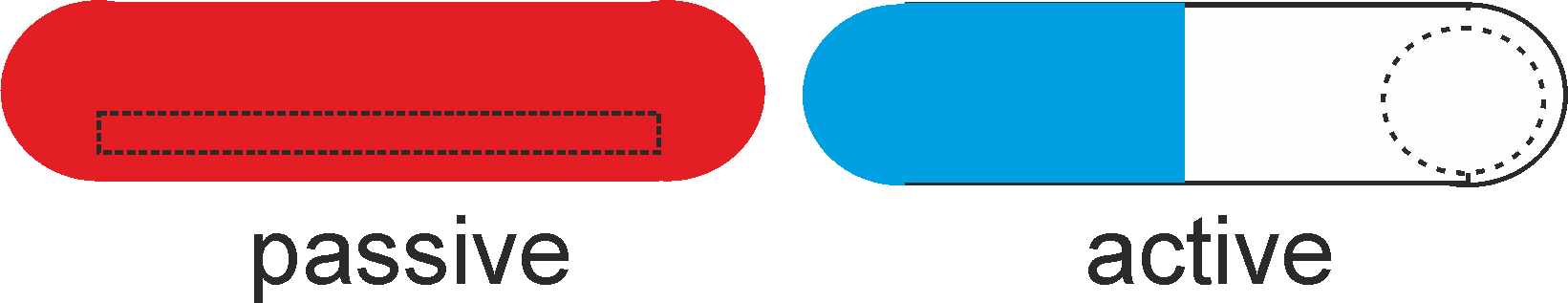}
        \caption{Sketch of the two types of particles. The position of the internal weights is indicated in a side view by dashed lines. Active capsules drift towards the white cap when excited. \label{fig:ParticleSketch}}
\end{figure}

The shells of all particles are commercial pharmaceutical  spherocylindrical capsules of 21 mm length and 7 mm diameter (Figure~\ref{fig:ParticleSketch}).
The artificial active particles contain a copper-plated lead bullet (air rifle ammunition) of 0.48 g mass which is asymmetrically glued inside the capsule at one end. The unequal mass distribution is the basis for the self propulsion on the shaken plane. Since this bullet has a smaller diameter than the capsule, it breaks the rotational symmetry of the particle so that sideways rolling of the capsule is inhibited. Active capsules have two colors, white and blue, in order to distinguish their heavier head (white) from the lighter tail (blue) in image processing. 

Passive capsules were prepared with same size and total mass as the active ones, but there are two fundamental differences to distinguish them from the active ones: First, the passive particles contain a thin aluminum bar of 15 mm length which is glued symmetrically along the major axis of the capsules. Their motion is symmetric with respect to the long axis of the capsule.
Second, the passive particles are mono colored (red). They appear symmetric in image processing.


\subsection{Particle excitation and observation}

\begin{figure}[htbp]

        \includegraphics[width=0.485\columnwidth]{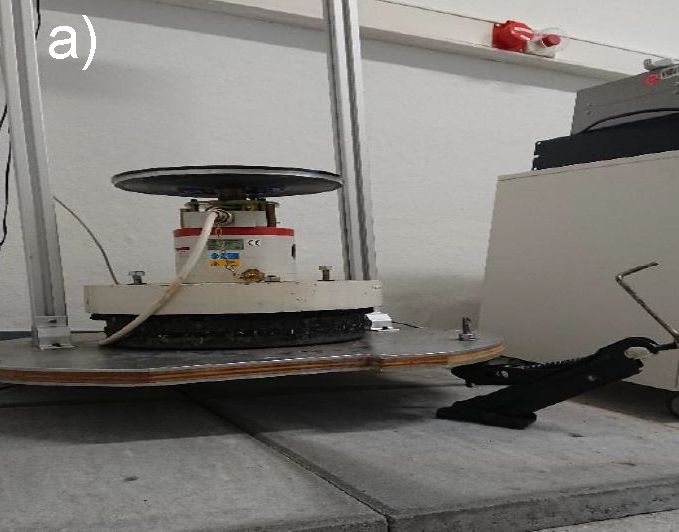}\hfill
        \includegraphics[width=0.5        \columnwidth]{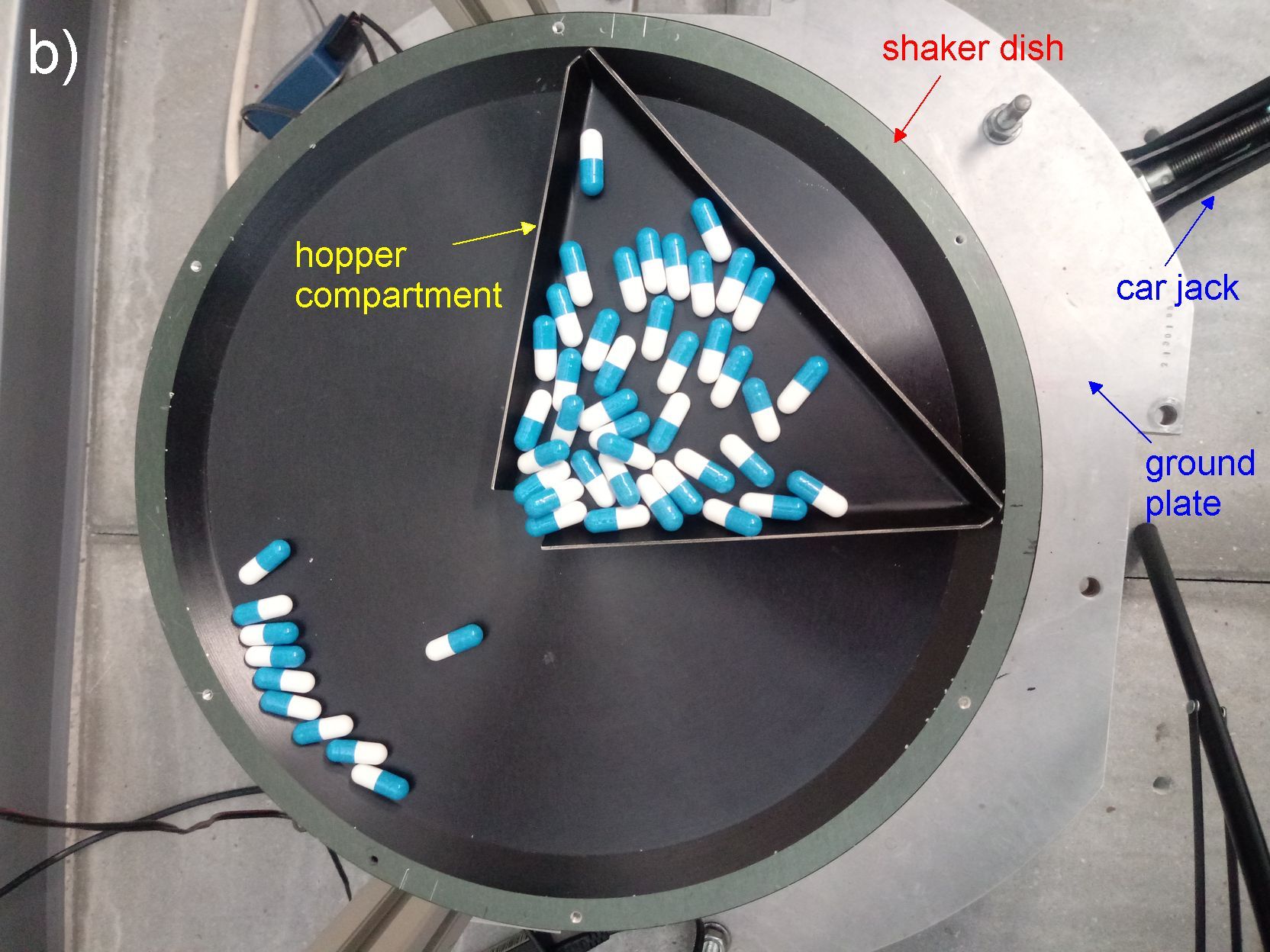}
        \caption{a) Side view of the experimental setup with shaker and car jack, the camera above is not shown.
        b) The dish with the hopper compartment: Capsules are initially placed on a triangular metal sheet mounted on the shaking dish. Its edges are bent upward. Two of these edges leave a gap at one side where the particles can exit. The motivation for the particles to pass this orifice is created by a slight tilt of the complete setup (up to $\approx 5^\circ$), using the car jack.
        \label{fig:1}}
\end{figure}

The particles are excited by vertical vibrations of a shaker dish with flat bottom and a slightly elevated border in order to prevent them from leaving the dish. Figure \ref{fig:1}a shows the side view of the experimental setup, without the cameras.
The shaker is driven by a sinusoidal wave form, whose amplitude and frequencies can be adjusted in a custom-made Labview program.
In order to provide an appropriate excitation for the particles, we adjusted the acceleration to values slightly larger than the gravitational acceleration ($g=9.81$~m/s$^2$) to 1.4~$g$, 1.6~$g$ and 1.8~$g$. The corresponding vibration amplitudes were in the range of a few dozen micrometers, the excitation frequencies in the range between 50 and 100 Hz.
Below the above given minimum excitation in our experiment, the motion of the particles is very slow. Above 1.8 $g$, the excitation becomes too strong, the particles begin to jump, to stand vertically frequently and topple over in the tilted hopper, and to leave the horizontal plate.

In order to observe the behaviour of the particles near a bottleneck we manufactured trianglular plates with different geometries serving as sample containers (artificial 2D 'hopper'). Figure \ref{fig:1}b shows the top view of the shaker with the hopper compartment filled with particles.
This metal sheet provides the confinement in our experiment. Seen from above, this sheet has funnel shape, its three edges are bent upward to serve as barriers for the particles. At one side where two edges approach each other, they leave a gap which serves as the bottleneck the particles have to pass.

To motivate the particles to move through the gap, the whole setup including the shaker and cameras is slightly tilted. We use a car jack to adjust angles of a few degrees with an accuracy of approximately 0.15$^\circ$.
The desired slopes in different experiments were between 2.5$^\circ$
and
4$^\circ$. At larger tilt angles, particles tend to jump and show a tendency to rise with one end above neighbors.
At angles smaller than 2$^\circ$, there is no noticeable trend of the capsules to move downward within the experimental time of approximate one minute.

\begin{figure}[htbp]

        \includegraphics[width=0.46\columnwidth]{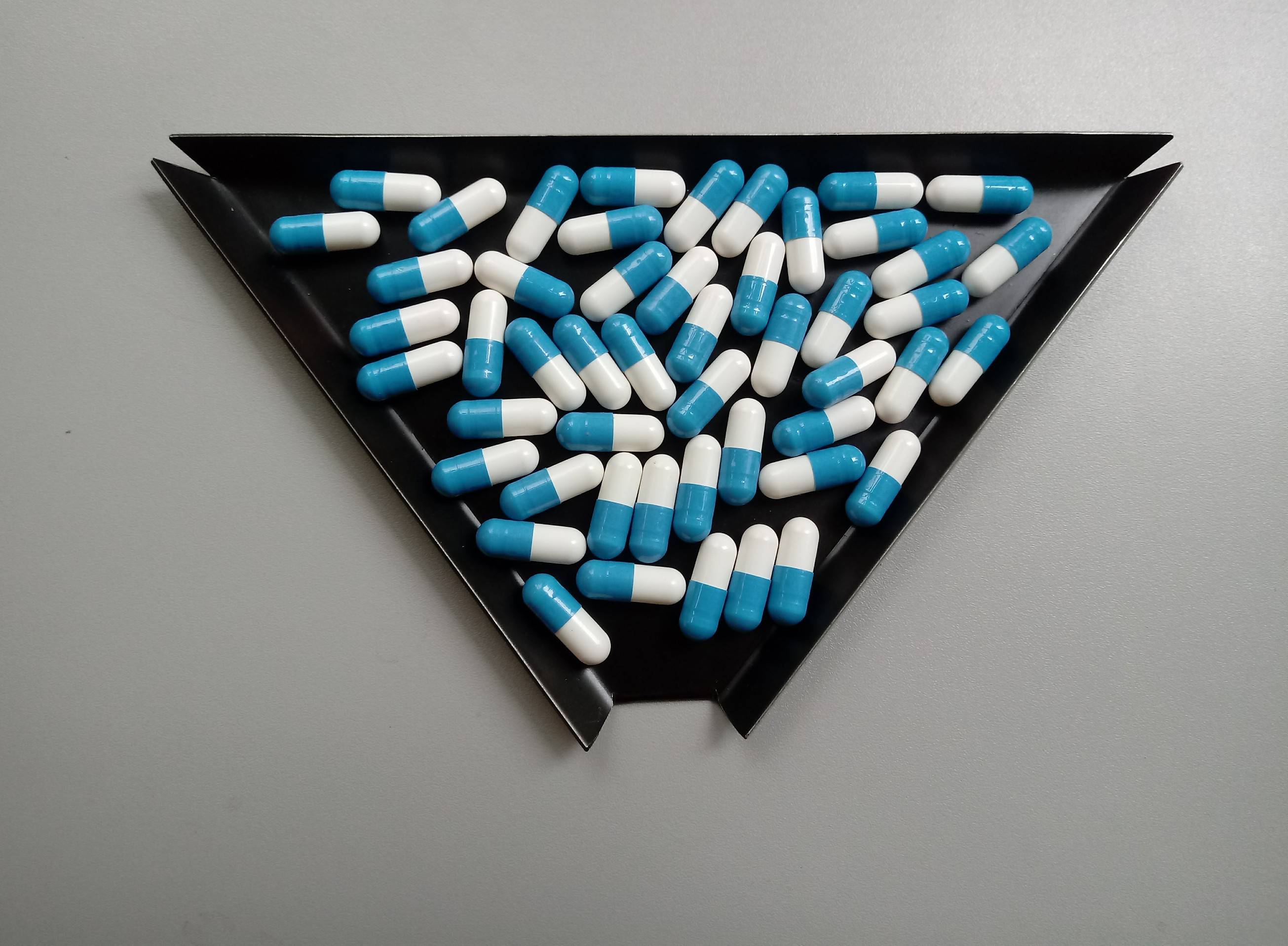}
       \includegraphics[width=0.45\columnwidth]{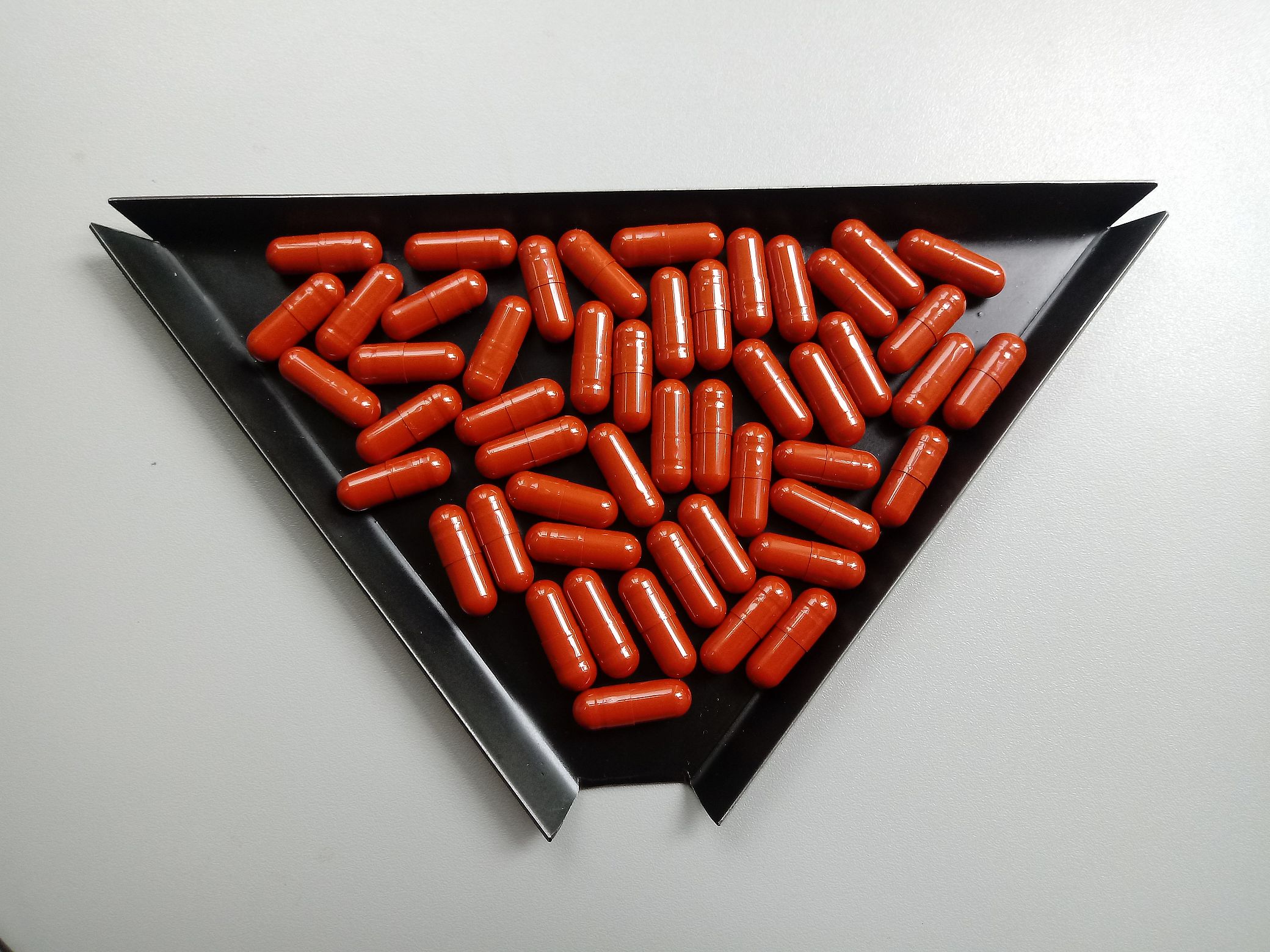}

        \caption{Active (left) and passive (right) particles in the initial state after filling.\label{fig:4}}
\end{figure}

This compartment is fixed onto the vibrating dish, with a few millimeters height gap between them. 
Then, the particles leaving the hopper through the gap drop down onto the dish and do not block the orifice after they passed it. The top edge of the compartment is 190 mm wide, its vertical distance to the orifice is 110 mm. The opening sizes were chosen as 15~mm, 18~mm and 21~mm in individual experiments.
Figure \ref{fig:4} shows the top view of the compartment initially containing the particles. The direction of the tilt towards the orifice breaks the symmetry for particle motion in the plane and defines the mean preferential direction of motion of both active and passive particles.

The particles were initially filled into the compartment while the shaker was at rest. The two images in Fig.~\ref{fig:4} show typical initial configurations of the compartment filled with active and passive capsules. Then, the shaker was switched on and the emptying of the silo during excitation was recorded with optical video imaging.
In order to characterize the individual particles, we have additionally performed an analysis of single capsules moving on a vibrating horizontal plate with the same excitation parameters but zero tilt. These
measurements show the qualitatively different kinematic behaviour of both species.

The collective bottleneck passage experiments were recorded with a Canon EOS customer camera at a frame rate of 60 fps with a resolution of 4 pixels/mm. The setup was illuminated with diffuse light in order to avoid reflexes at the particle surfaces which are undesired in image processing. The diffusion experiments with single capsules were performed with a XiaoYi 4K camera with a resolution of 26 pixel/mm at 100 fps.
The same camera and frame rate were used to record close-ups of the orifice 
to determine the time intervals between the passages of active particles.
The typical duration of a discharge experiment was of the order of one minute. The video shown in the supplemental information shows downgraded footage of a typical experiment.

\subsection{Image processing and analysis}

In order to obtain the trajectories of the capsules, one first needs to detect the particles by means of image processing and analysis. The detection of single particles in a video frame can be performed by separation of colors in the image. For a passive particle, its center of mass and orientation are then easily extracted using MATLAB as segmented region properties. For an active particle, one can find centroids of colored (blue and white) regions and the orientation of the vector between them.
Moreover, if particles in the image are lying horizontally on the plate and do not overlap in the camera view, simultaneous detection of multiple capsules is possible. The regions corresponding to each single passive particle are easily separated from each other, so the detection of multiple particles is easy in this case.

\begin{figure}[htbp]

      \includegraphics[width=0.43\columnwidth]{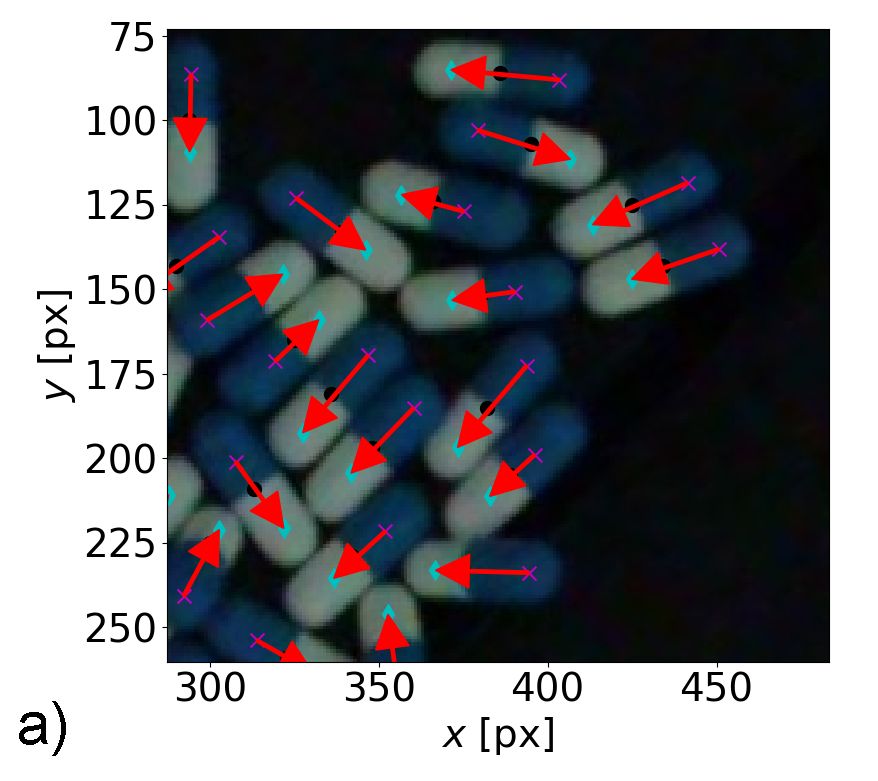}
      \includegraphics[width=0.55\columnwidth]{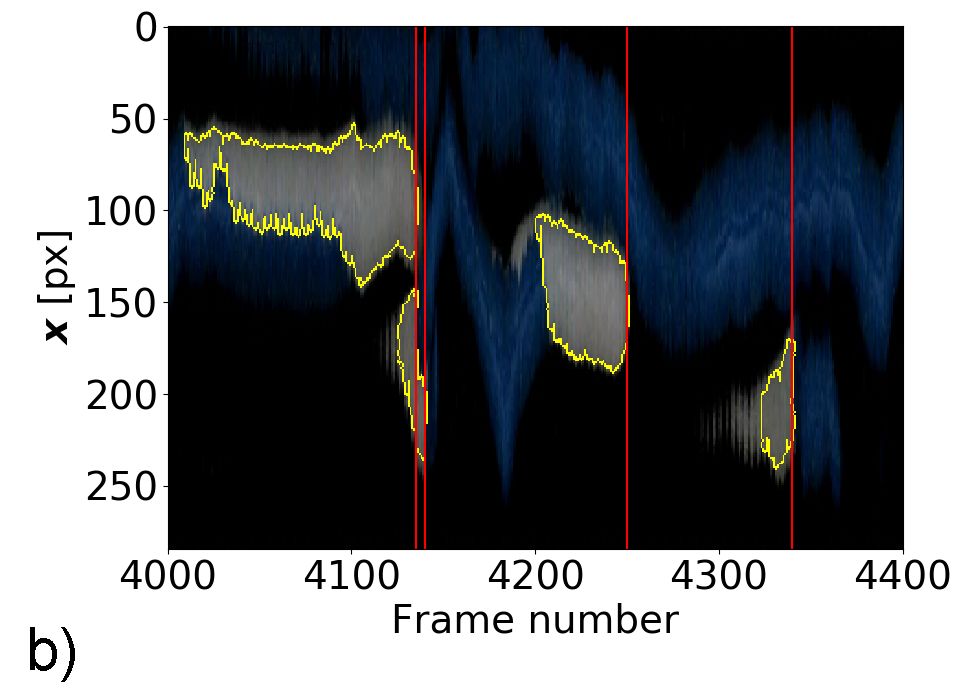}
      \caption{a) Example of the detection of active particles. Red arrows show the detected directions of the capsules. b) Extraction of passage times from the space-time plot sampled at the orifice. Red lines correspond to frames where a passage  occured. \label{fig:Detection}}
\end{figure}

In order to reliably find the positions and orientations of multiple active particles, one has to perform an additional step to match image regions corresponding to the blue and white capsule parts of one and the same capsule. The optimal assignment problem can be solved so that all vector lengths between matched blue and white regions' centroids correspond to the parameters of real particles. This avoids a situation where parts of different capsules are matched.
Figure~\ref{fig:Detection}a) demonstrates the result of the detection of active particles, zoomed in from one exemplary video frame.

After detecting particles on each video frame, the Crocker-Grier algorithm \cite{Crocker1996} is applied to find their trajectories. While tracking works well for a single particle as well as multiple non-overlapping particles, it is problematic closer to the orifice, where capsules often jump, assume tilted positions or even partially overlap. 
This is true especially for higher  excitation amplitudes and smaller orifice sizes.

In the situations where particle tracking is problematic, one can adopt another method which allows to extract the times when a particle leaves the silo (passage times). A similar technique was previously used for the extraction of data in experiments with sheep and humans \cite{Zuriguel2014, Zuriguel2016, Garcimartin2016} and soft grains \cite{Harth2020}. Here, from each video frame a thin image slice from a line across the outlet is sampled. If the frame rate is sufficiently high, one obtains a long space-time image with blue and white regions corresponding to each capsule which passes the orifice by stacking these slices. It suffices to focus on the more easily detectable white capsule parts, while they are oriented back from the orifice for the most particles leaving the silo. Moreover, it is mentioned in Ref.~ \cite{Zuriguel2016} that the exact part of the object that is chosen as a reference to measure the passage time does not substantially affect the statistics. Figure~\ref{fig:Detection}b shows a short cutout from this image which corresponds to 400 frames (4 seconds).
By finding the rightmost (latest) borders of the segmented white regions, marked by red lines in Fig.~\ref{fig:Detection}b), one gets the frame numbers and hence exact times when capsules pass the outlet. From these data we obtain the number of particles remaining in the silo at each instant of time during the discharge, as well as the times between subsequent particles leaving it. Additionally, one can extract the  orientations of the capsules.

The identification of passive capsules can be done analogously. As mentioned before, these capsules are completely red. In order to detect red particles, the red channel is extracted with MATLAB and then subtracted from the specified background
(same background subtraction for all images). Finally, we analyze 
the properties of the detected objects such as the centers of mass, positions, orientations and areas for each frame. 
As the passive particles show practically no overlaps and do not leave the hopper plane, this simple particle detection is sufficient to access their egress and orientation behaviour.

\section{Results}
\subsection{Single particle kinematics}
\label{Sec:Single}
\begin{figure}[htbp]
      \includegraphics[width=0.49\columnwidth]{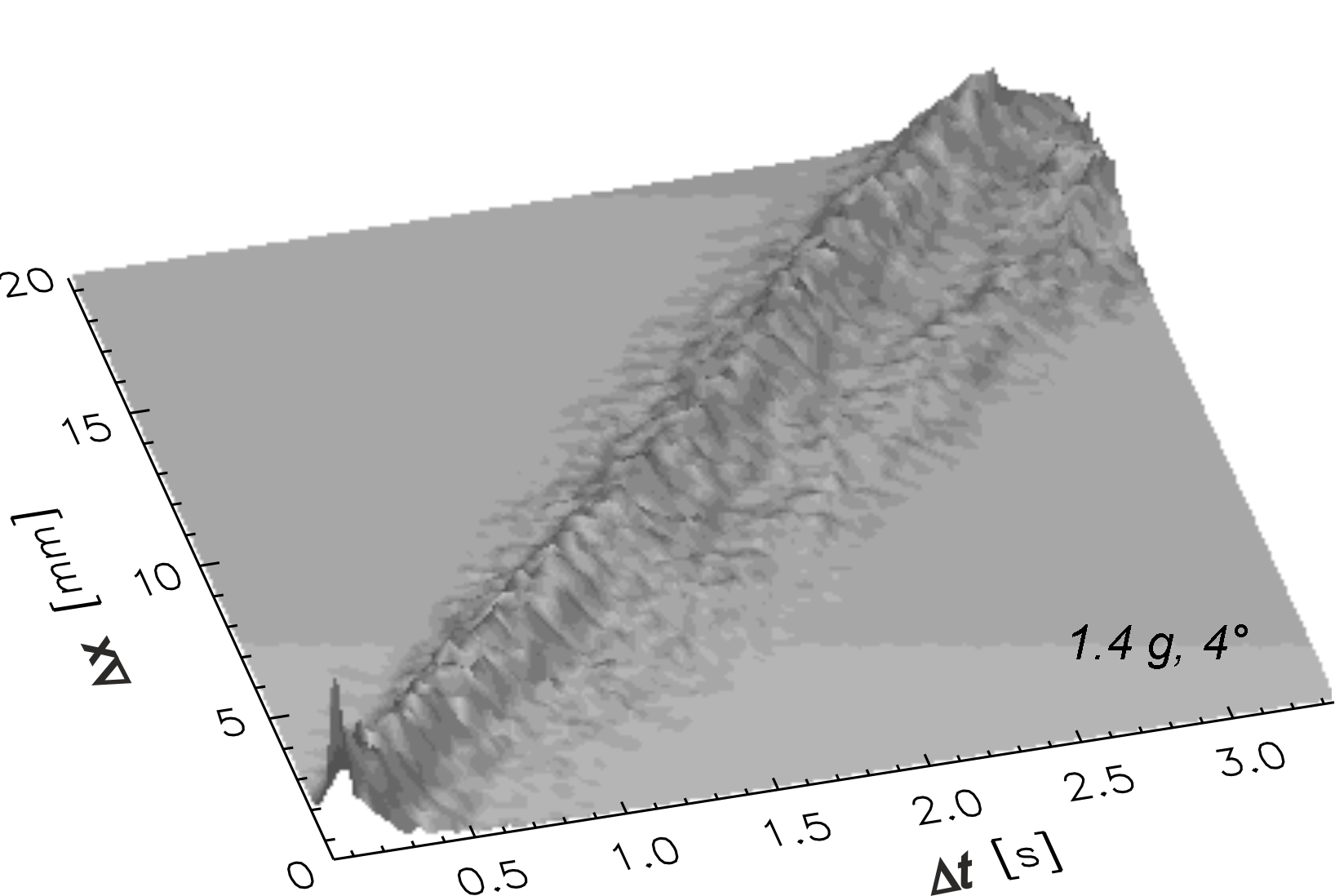}
      \includegraphics[width=0.49\columnwidth]{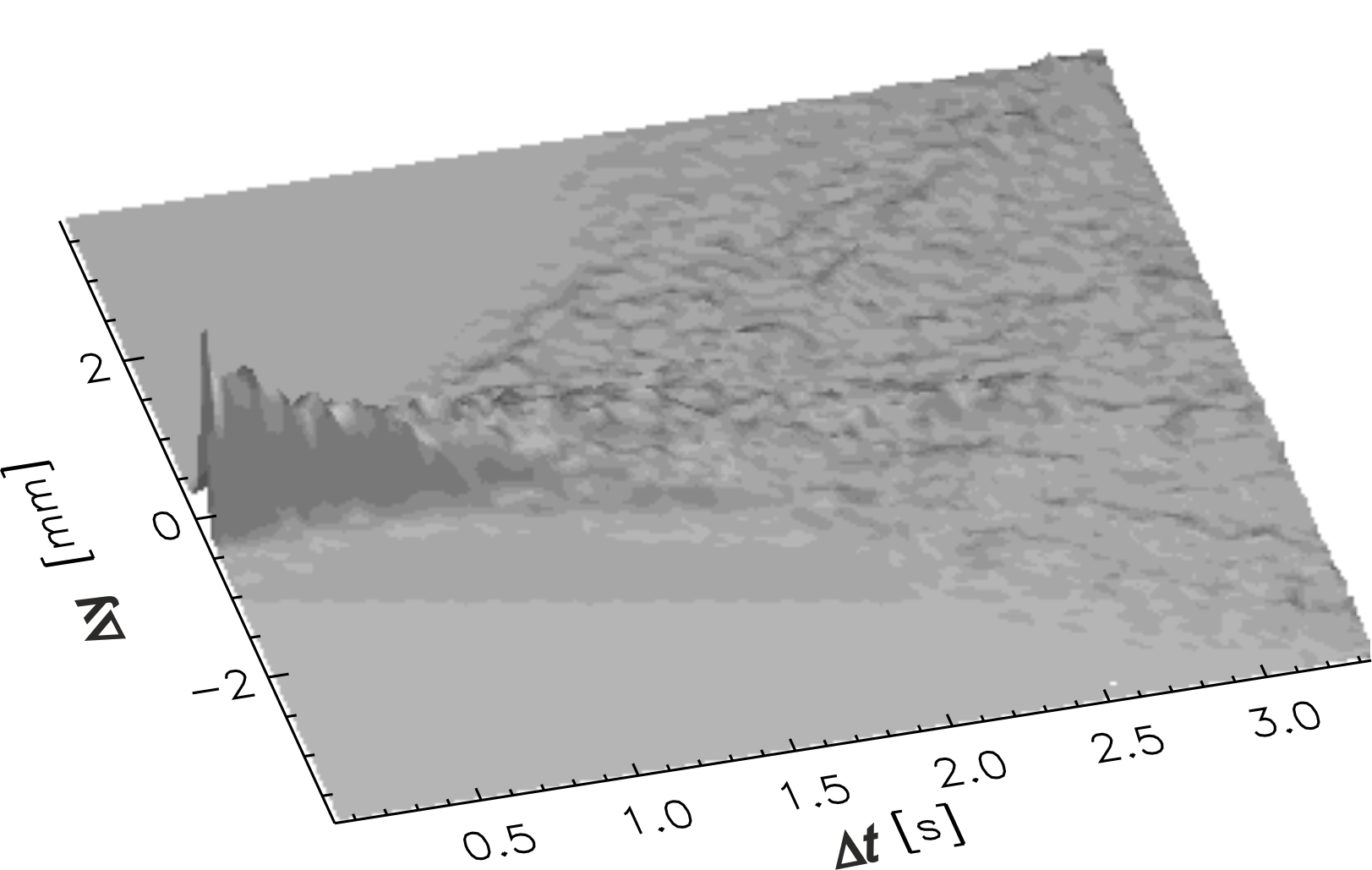}
      \caption{Distribution of displacements between two video frames separated by the delay $\Delta t$ for an active capsule. The left image shows the propagation along the momentary capsule axis, the right image shows the same in the direction perpendicular to that axis. The excitation strength was 1.4 $g$. More such plots can be found in the Appendix A \label{fig:hist}}
\end{figure}

For the determination of the single particle motion, trajectories of isolated grains were recorded and their mean square displacements (MSD) and
mean square rotational displacements were measured.
Figure \ref{fig:hist} shows the evolution of the distribution of displacements with progressing delay time for isolated active particles on a horizontal vibrating plate \cite{SI1}.
It is clearly evident that there is a drift along the capsule axis, and a slight broadening of the distribution by additional random, diffusive contributions. In contrast, the distribution of  displacements perpendicular to the capsule axis remains symmetric with respect to the two sides. It is a pure  diffusion process without drift.
The mean value of this distribution remains close to zero.

Figure \ref{fig:msd} shows typical mean square displacements for the active particles. The displacement along the capsule axis is nearly two orders of magnitude larger than that in perpendicular direction. Moreover, the MSD grows with the square of $\Delta t$. Here, we have chosen a short period of 2 seconds where the change of the direction of motion by rotational diffusion is still small. The displacements refer to the initial capsule orientation, which can change by a few degrees within the evaluation period $\Delta t$. The diffusion perpendicular to the capsule axis is mainly caused by such fluctuations of the long axis direction, therefore it is also proportional to $\Delta t^2$.

\begin{figure}[htbp]
\centering
      \includegraphics[width=0.8\columnwidth]{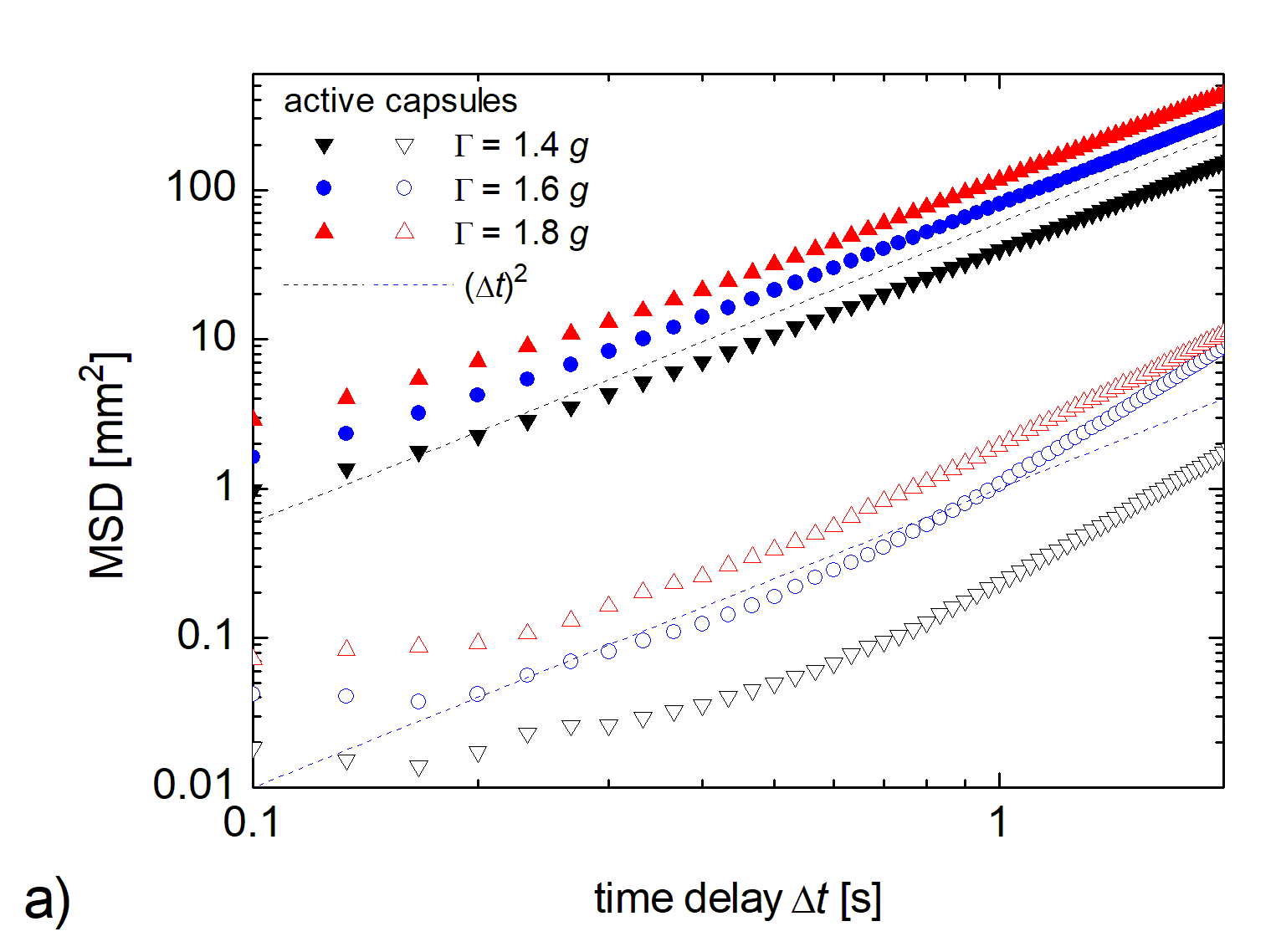}
      \includegraphics[width=0.8\columnwidth]{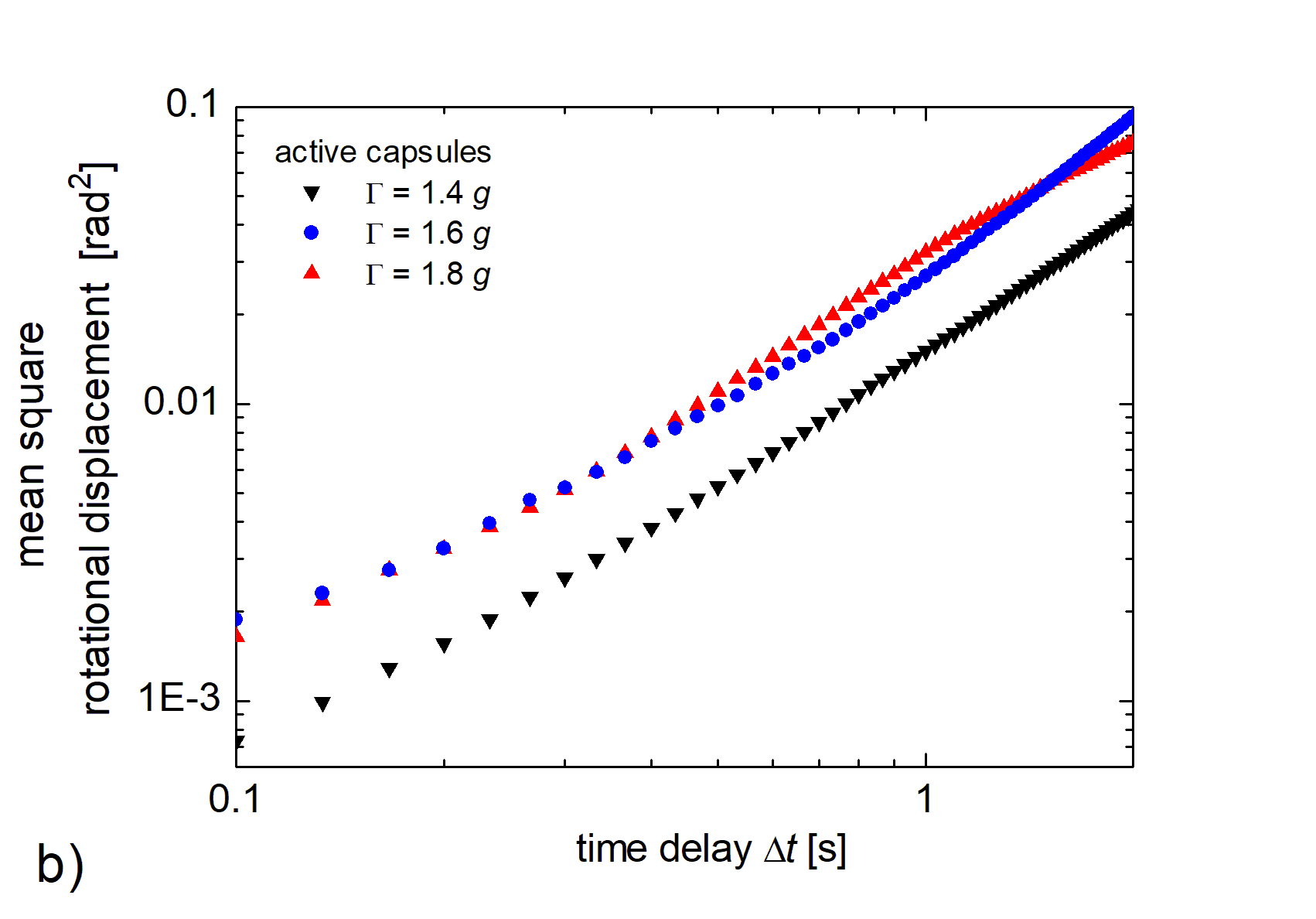}
      \caption{a) Mean square displacements of active capsules along the capsule axis (filled symbols) and perpendicular to it (open symbols).  b) Mean square angular displacements of active capsules. Colors refer to excitations with maximum accelerations 1.8 $g$ (red), 1.6 $g$ (blue) and 1.4 $g$ (black)\label{fig:msd}}
\end{figure}

Particle trajectories keep their current directions for a few seconds, over some persistence length of the order of the capsule length (see \cite{SI1}). On long time scales ($\Delta t\gg 2$ s), purely diffusive behavior takes over, which is well known for active grains \cite{Howse2007}.

\begin{figure}[htbp]
\centering
      \includegraphics[width=0.8\columnwidth]{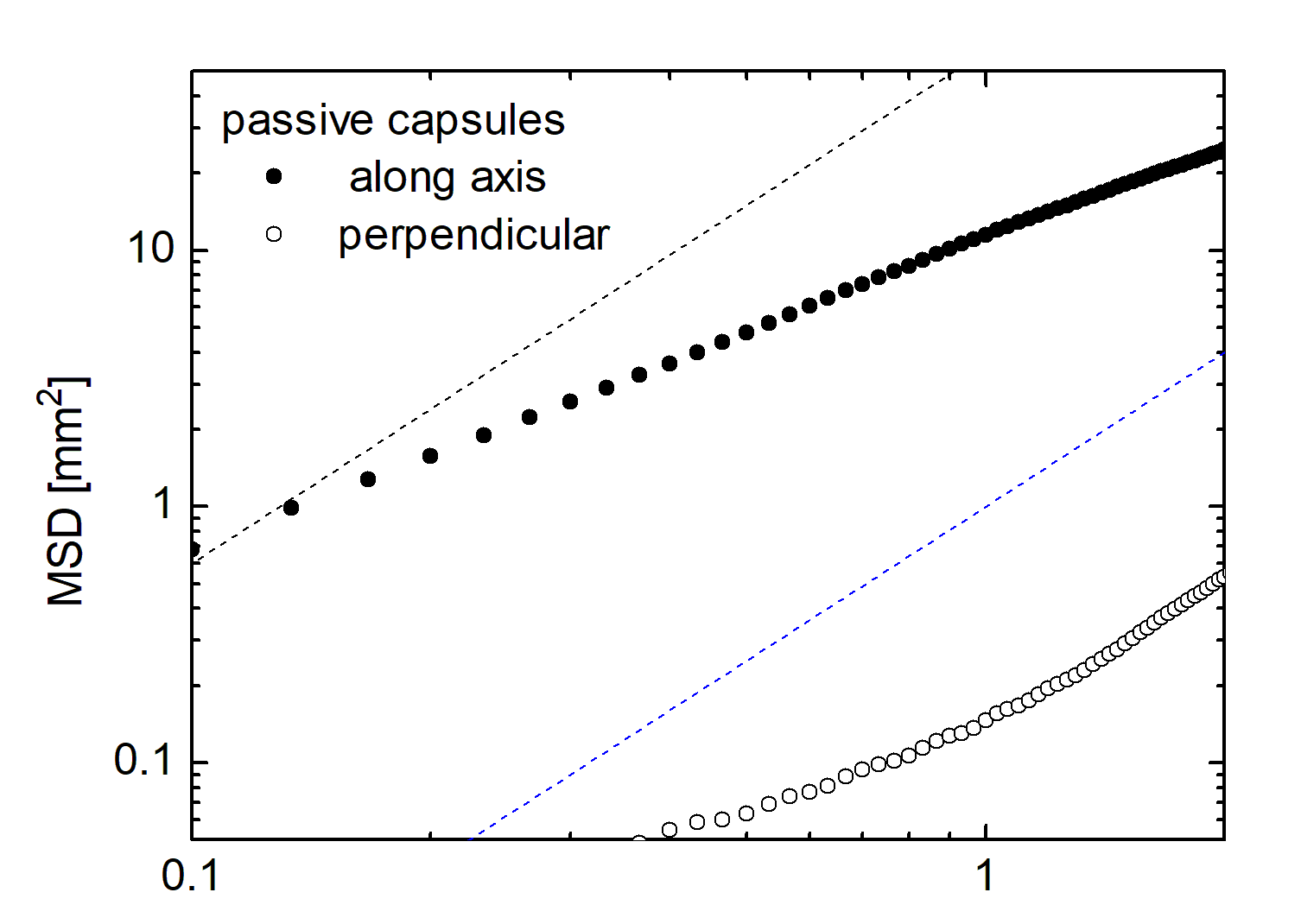}
      \caption{Mean square displacements of passive capsules along the capsule axis (filled symbols) and perpendicular to it (open symbols). The excitation strength was 1.6 $g$ . Dashed lines are the same as in Fig.~\ref{fig:msd}a, in order to provide a comparison to the active capsules.
      \label{fig:MSDpass}}
\end{figure}

For the passive particles, the distributions of both the displacements along and perpendicular to the particle axis are symmetric with respect to positive and negative displacements. The passive particles diffuse much more slowly than the active ones, as seen in Fig.~\ref{fig:MSDpass}. Along their axis direction, they diffuse more than one order of magnitude faster than perpendicular to it. The MSD curves parallel to the long axis grows linearly with the delay time $\Delta t$. The perpendicular diffusion has a stronger dependence on $\Delta t$, since diffusion parallel to the axis is 'mixed in' due to rotational diffusion (see paragraph on active particles). Typical distributions of displacements in dependence of $\Delta t$ are shown in the Appendix A in Fig.~\ref{fig:A1}.
This
appendix also shows more results for active capsules at different excitation strengths. The essential message from the single-particle measurements is that we have qualitatively distinct behaviour of both types of capsules that are quantitatively reproducible.
On the basis of these results, we investigate now how the differences in the individual particle motions influence the collective behavior in the bottleneck passage experiment.

We conclude from these measurements that the asymmetrically loaded capsules are useful models for active matter, and a comparison of their individual and collective behavior with that of the symmetrically loaded, passive particles will yield useful insight on the role of activity in several experimental situations like passing bottlenecks, self-organization of collective motion (swarms), sedimentation, and others.

\subsection{Collective passage through narrow bottlenecks}

\begin{figure}[htbp]
\centering
        \includegraphics[width=0.8\columnwidth]{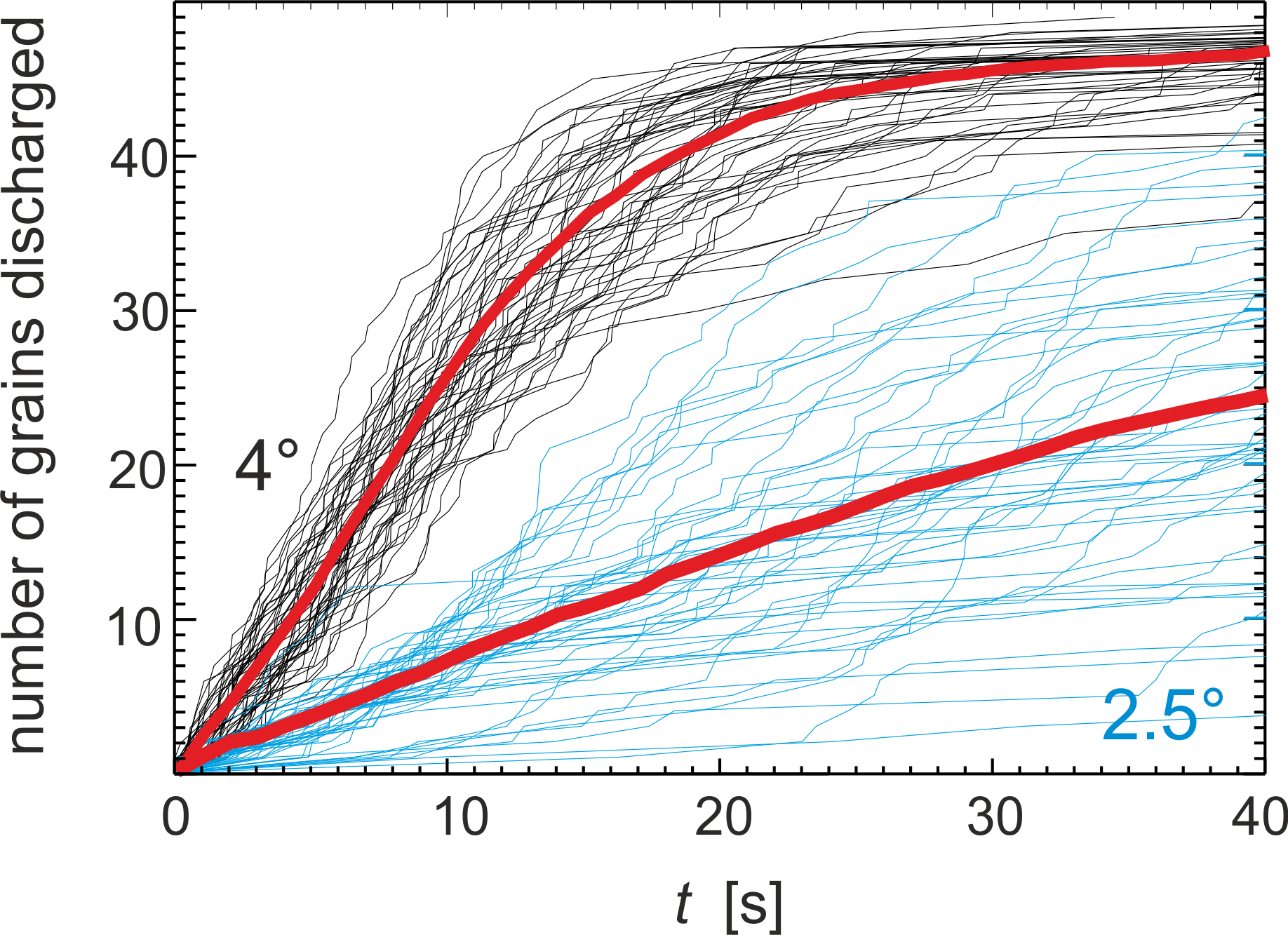}
        \caption{Number of active capsules that have left the container as a function of time, starting with the exit of the first capsule. All data correspond to $d=18$~mm and $\Gamma=1.6~g$. The black curves represent $\theta = 4^\circ$ tilt, the blue ones 2.5$^\circ$. The red graphs reflect the respective averages (mean number of particles that have left the exit after time $t$).   \label{fig:flowplot}}
\end{figure}

\begin{figure}[htbp]
\centering
        \includegraphics[width=0.8\columnwidth]{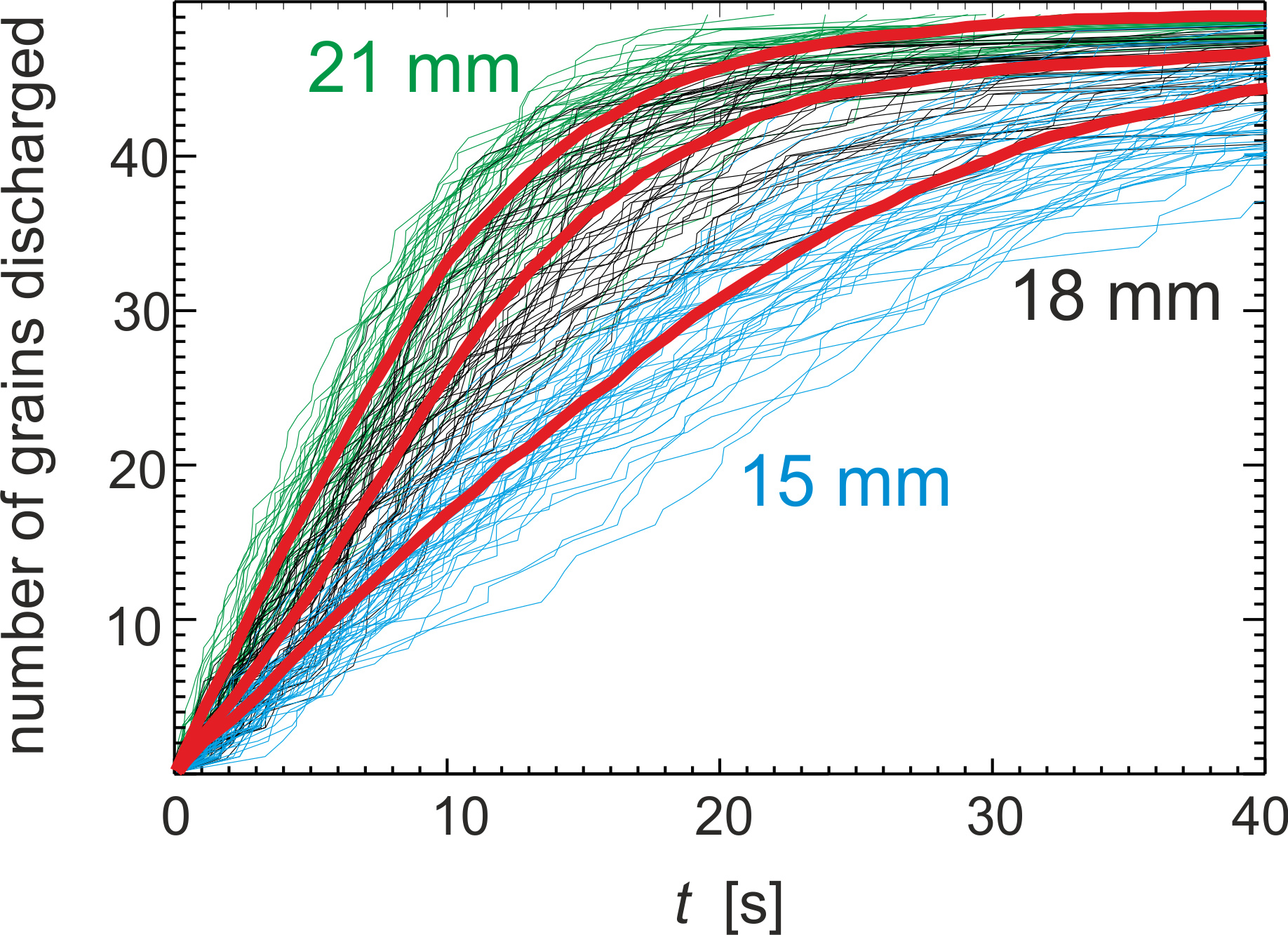}
        \caption{Number of active capsules that have left the container as a function of time, starting with the exit of the first capsule. All data correspond to $\theta=4^\circ$ and $\Gamma=1.6~g$. The black curves represent individual runs with $d=18$~mm, green curves 21~mm, blue curves 15~mm. Red graphs reflect the averages (mean number of particles that have left the exit after time $t$). The initial rates $\dot N$ are 1.7/s, 2.55/s, and 3.6/s, respectively, for the orifice sizes 15~mm, 18~mm, and 21~mm.  \label{fig:flowplot2}}
\end{figure}

First, we checked how the fill level of our container affects the discharge characteristics. Figure \ref{fig:flowplot} shows two examples of the outflow
characteristics of active grains. We selected the $d=18$~mm orifice data at 1.6~$g$ excitation strength and varied the tilt angle. It is seen that individual discharges vary statistically around an average value (red curves) for each tilt angle. The discharge curves are on average rather linear as long as enough grains remain in the container. This means that the mean outflow rate is
roughly constant until about 2/3 of the initial filling has left the container. The last $\approx 15$ particles discharge slightly slower than the first ones. In particular, the outflow rate drops significantly when less than 10 capsules remain in the container. Some of these remaining grains are stuck head-on at the side walls in positions that they keep for a long time when
there are no other capsules colliding with them.

\begin{figure}[htbp]
\centering
        \includegraphics[width=0.8\columnwidth]{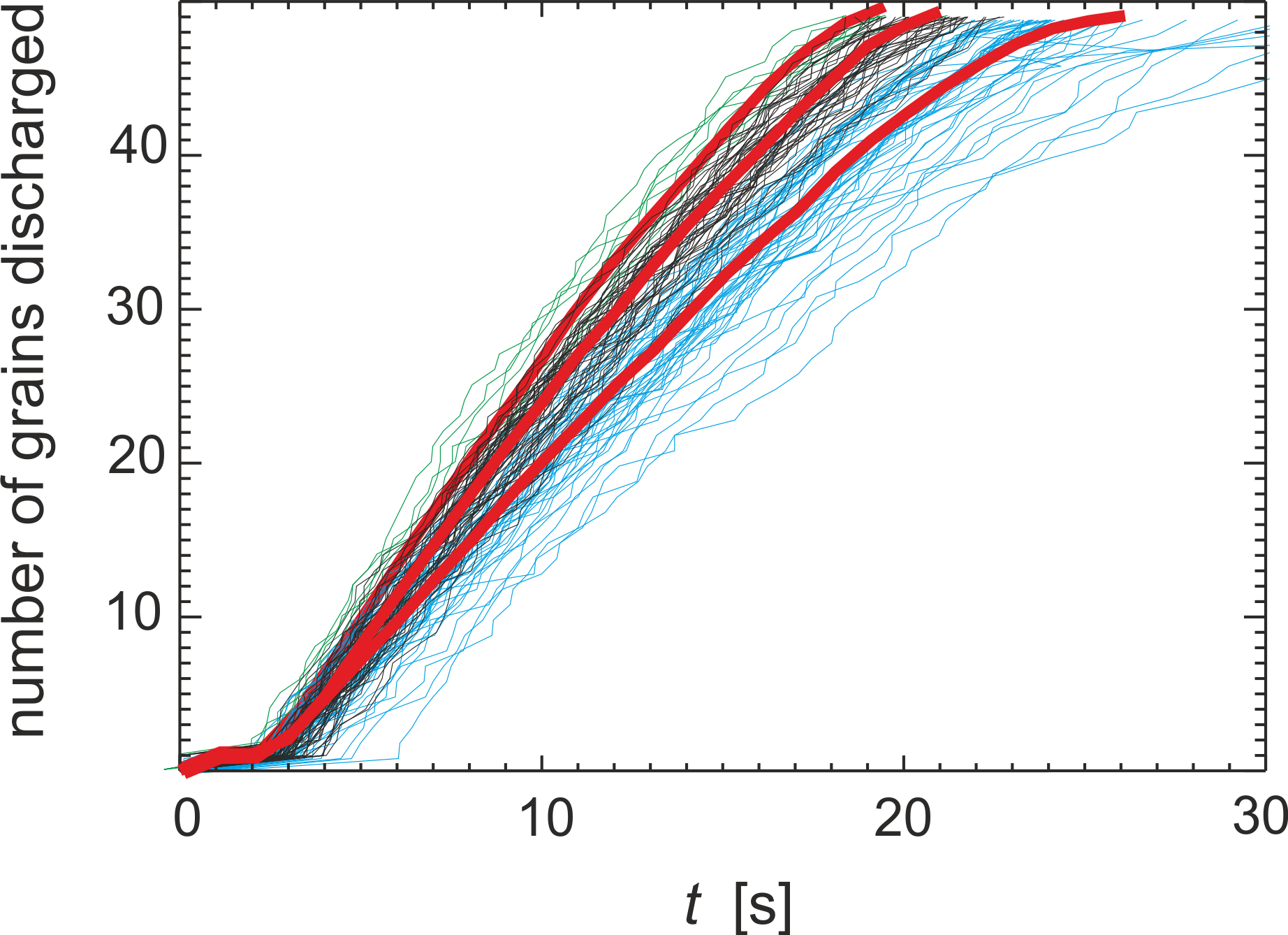}
        \caption{Number of passive capsules that have left the container as a function of time, starting with the exit of the first capsule. All data correspond to $\theta=4^\circ$ and $\Gamma=1.6~g$. The black curves represent individual runs with $d=18$~mm, green curves 21~mm, blue curves 15~mm. Red graphs reflect the averages (mean number of particles that have left the exit after time $t$). The initial rates $\dot N$ are 2.5/s, 2.9/s, and 3.4/s, respectively, for the orifice sizes 15~mm, 18~mm, and 21~mm  \label{fig:flowplotpass}}
\end{figure}

An interesting aspect is the dependence of the discharge rate on the orifice size. There are different predictions for pedestrians or animals passing a gate, where the rate $\dot N$ of individuals exiting is more or less linearly related to the gate width $d$, $\dot N\propto d^\beta$ with $\beta =1$. In contrast, the outflow rate of granular material, in reasonable approximation described by Beverloo's law,
goes with the exponent $\beta=3/2$ for 2D systems. This can be derived from a simple geometric arguments: the product of the orifice width and the velocity of particles passing the exit. When one assumes that the granular particles in a silo fall from an arch with approximate height $d$ above the orifice, the exit velocity is $\propto \sqrt{d}$, which yields $\beta=3/2$. In our system of passive particles (Fig.\ref{fig:flowplotpass}), we find rates of 2.5/s, 2.9/s and 3.4/s for the three orifice widths 15 mm, 18 mm and 21 mm, respectively, from  the linear slopes of the averaged graphs when the container is well-filled. This yields an exponent $\beta\approx 1$. This is a quite
reasonable result: The width of the orifice determines the number of particles passing per time, and they all pass with roughly the same velocity since they are not in free fall. They are subject to a continuous drift down the slope. The surprising result is found for the active particles. Here, the exponent is $\beta= 2.2\pm 0.2$ for the three mean discharge curves in Fig.~\ref{fig:flowplot2}. This behavior is qualitatively different from that of passive capsules.

A key feature of passages of living matter through bottlenecks is their
delay time between individuals passing the outlet \cite{Zuriguel2014}.
We extract this characteristics directly from our experiments. Each graph shown below is constructed from several independent discharge runs, the active capsule data, for example, from 50 individual runs each for each parameter combination. The delay times $\tau_i$ between two objects subsequently passing the exit are distributed according to a density $p(\tau)$ with
$\int_0^\infty   p(\tau) d\tau=1$. One can define a mean delay time $\bar\tau$ under the condition that the integral
$\bar\tau=\int_0^\infty \tau p(\tau) d\tau$ is finite, otherwise one needs to resort to
other measures describing the overall discharge characteristics.
The probability of delays being longer than a time $\tau$ is given by the complementary
cumulative distribution function (CDF)
\begin{equation}
G(\tau) = \int_\tau^\infty p(\tau') d\tau'    .
    \label{eq:taudef}
\end{equation}
with $G(0)=1$ and $G(\infty)=0$.
This cumulative distribution of experimental data is much easier to handle than $p(\tau)$. We have no theoretical prediction of the actual form of $G(\tau)$ but it turns out that the empirical ansatz
\begin{equation}
G(\tau) = \frac{1}{1+(\tau/\tau_0)^\gamma}
    \label{eq:tau2}
\end{equation}
with two adjustable parameters $\tau_0$ and $\gamma$ is well suited to describe all experimental data with desired accuracy.
The asymptotic tail of this distribution yields a power law dependence
$G \propto \tau^{-\gamma}$ which can be directly compared to data from
other studies \cite{Zuriguel2014}.
We note that the form of $p(\tau)$ is then not as simple,
\begin{equation}
    p(\tau) = \frac{\gamma}{\tau_0}\frac{(\tau/\tau_0)^{\gamma-1}}{[1+(\tau/\tau_0)^\gamma]^2}
    \label{eq:tau3}
\end{equation}
An advantage of the form of the cumulative function is that the integral
\begin{equation}
    \int_0^\infty G(\tau)d\tau = \int_0^\infty \tau p(\tau) d\tau = \bar\tau
    \label{eq:tau_mean}
\end{equation}
exists for $\gamma > 1$ and yields the analytical expression
\begin{equation}
  \bar\tau  = \frac{\pi}{\gamma \sin \pi/\gamma} \,\tau_0
\end{equation}
for the mean delay time $\bar\tau$ as a function of the fit parameters $\tau_0$ and $\gamma$. The very good agreement between the experimental $G(\tau)$ values and the empirical fit function Eq. (\ref{eq:tau2}) is demonstrated in Appendix B.

\begin{figure}[htbp]
\centerline{
      a)\includegraphics[width=0.58\columnwidth]{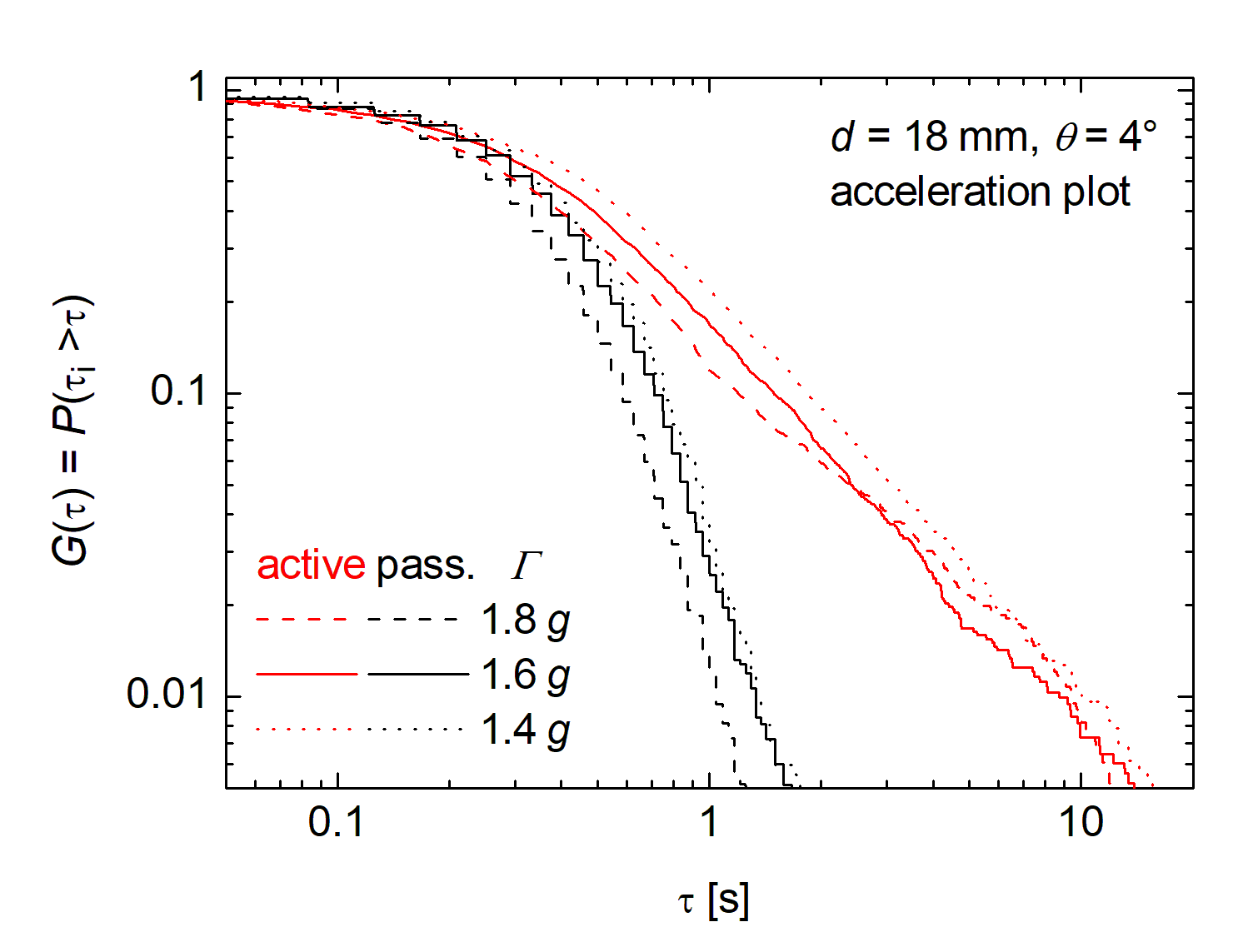}\hfill
      b)\includegraphics[width=0.39\columnwidth]{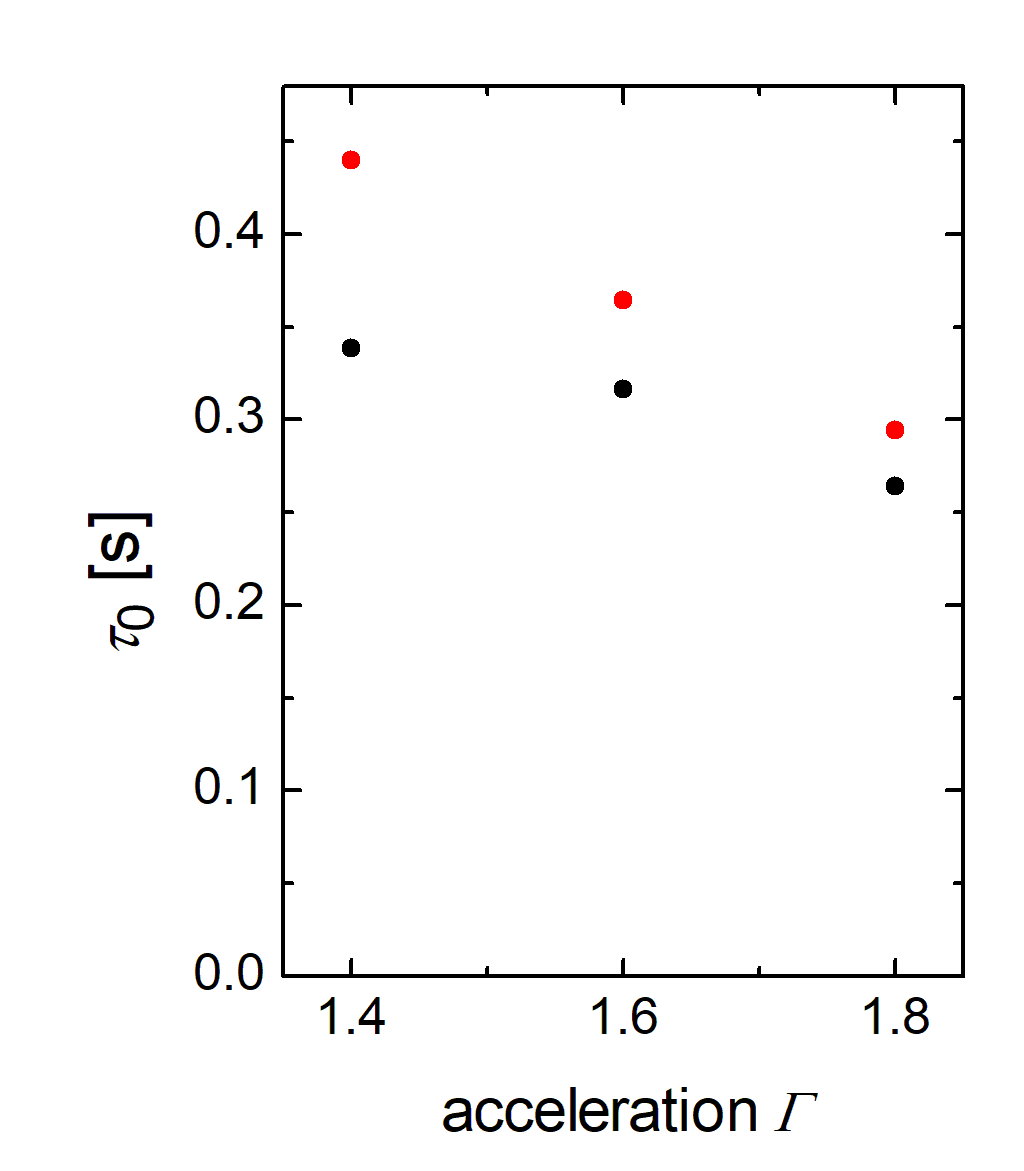}}
      \caption{a) Excitation strength dependence of $G(\tau)$ for active and passive capsules. The excitation strength is measured in terms of the maximum acceleration of the vibrating plate in units of the gravitational acceleration $g$. b) corresponding fit parameters $\tau_0$ (Eq.~(\ref{eq:tau2})).
      \label{fig:exc}}
\end{figure}

\begin{figure}[htbp]
\centerline{
      a)\includegraphics[width=0.6\columnwidth]{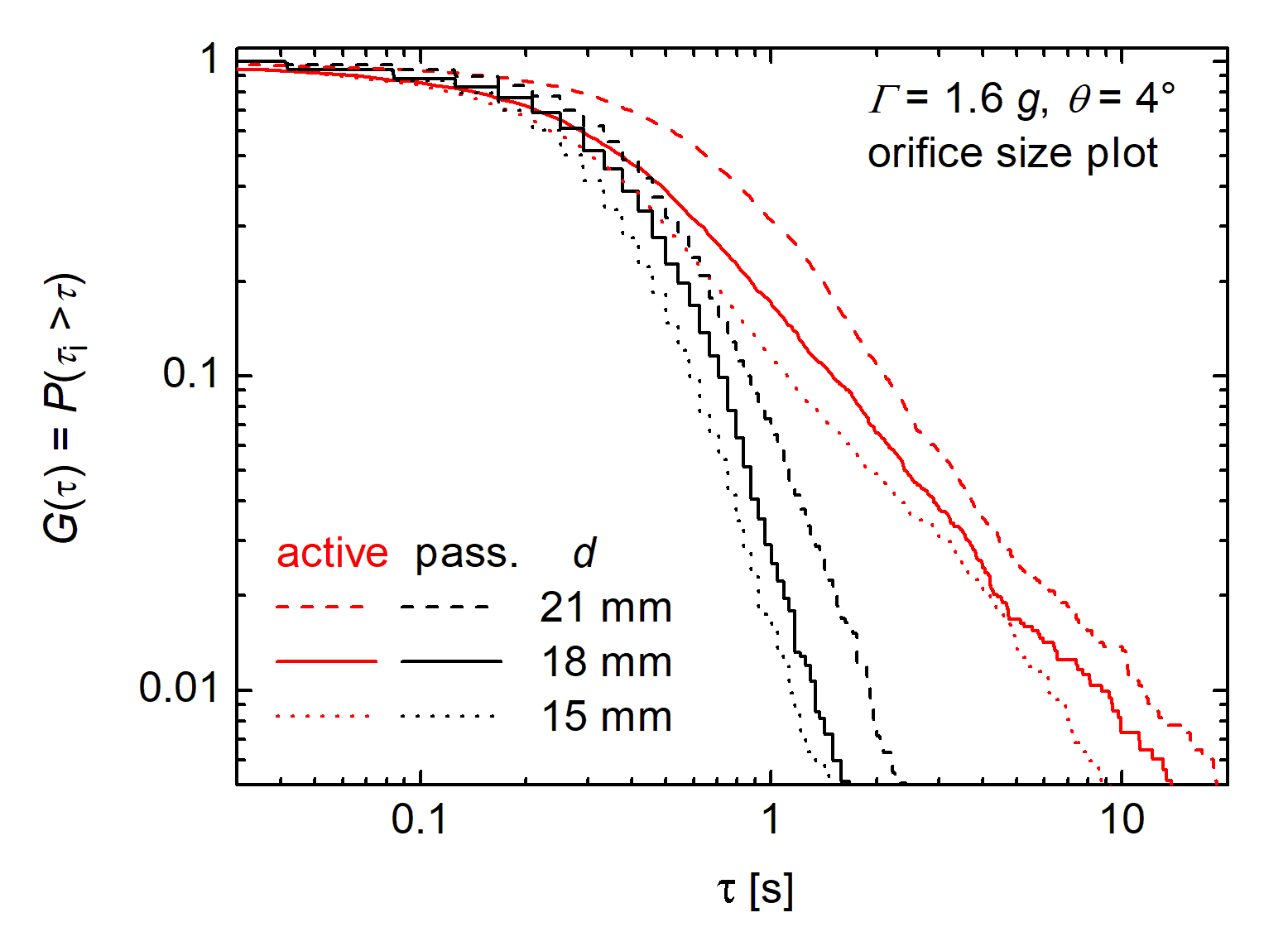}\hfill
      b)\includegraphics[width=0.3\columnwidth]{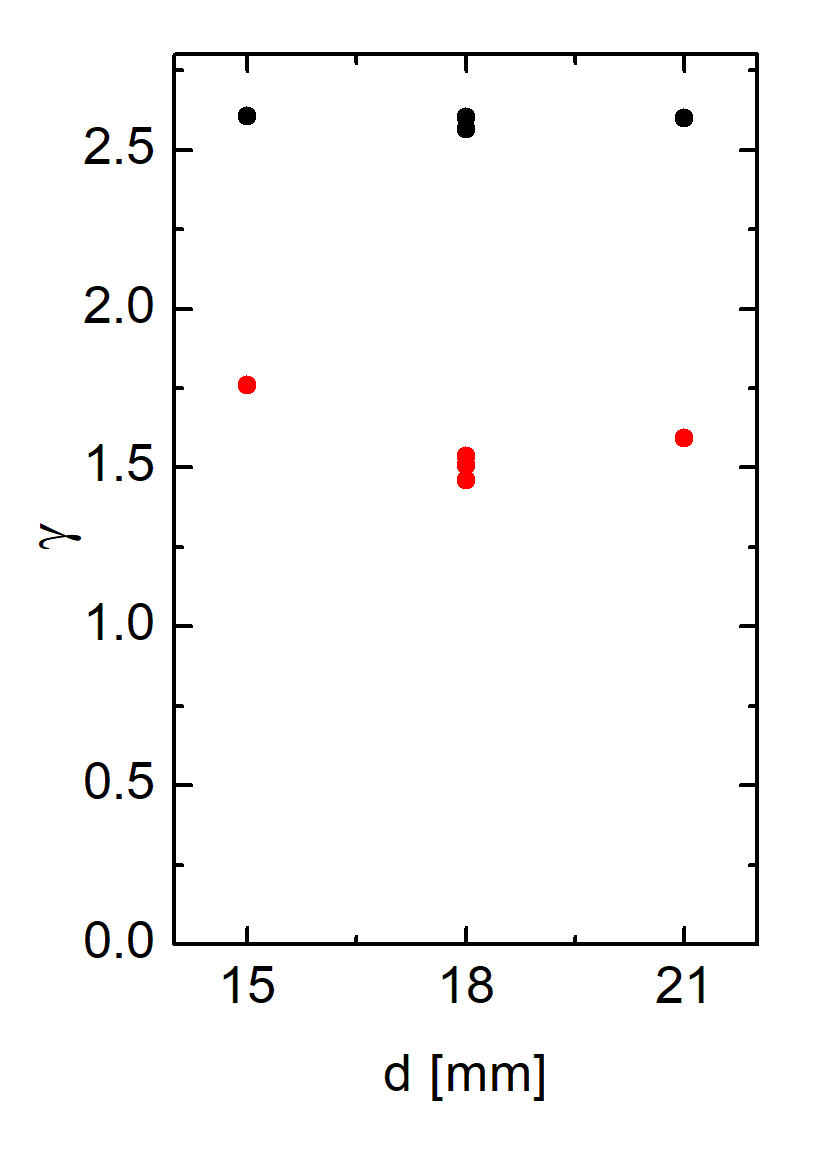}}
      \caption{a) Orifice size dependence of $G(\tau)$ for active and passive capsules. The excitation strength is measured in terms of the maximum acceleration of the vibrating plate in units of the gravitational acceleration $g$. b) corresponding fit parameters $\gamma$ (Eq.~(\ref{eq:tau2}))
      \label{fig:ori}}
\end{figure}

\begin{figure}[htbp]
\centering
      \includegraphics[width=0.8 \columnwidth]{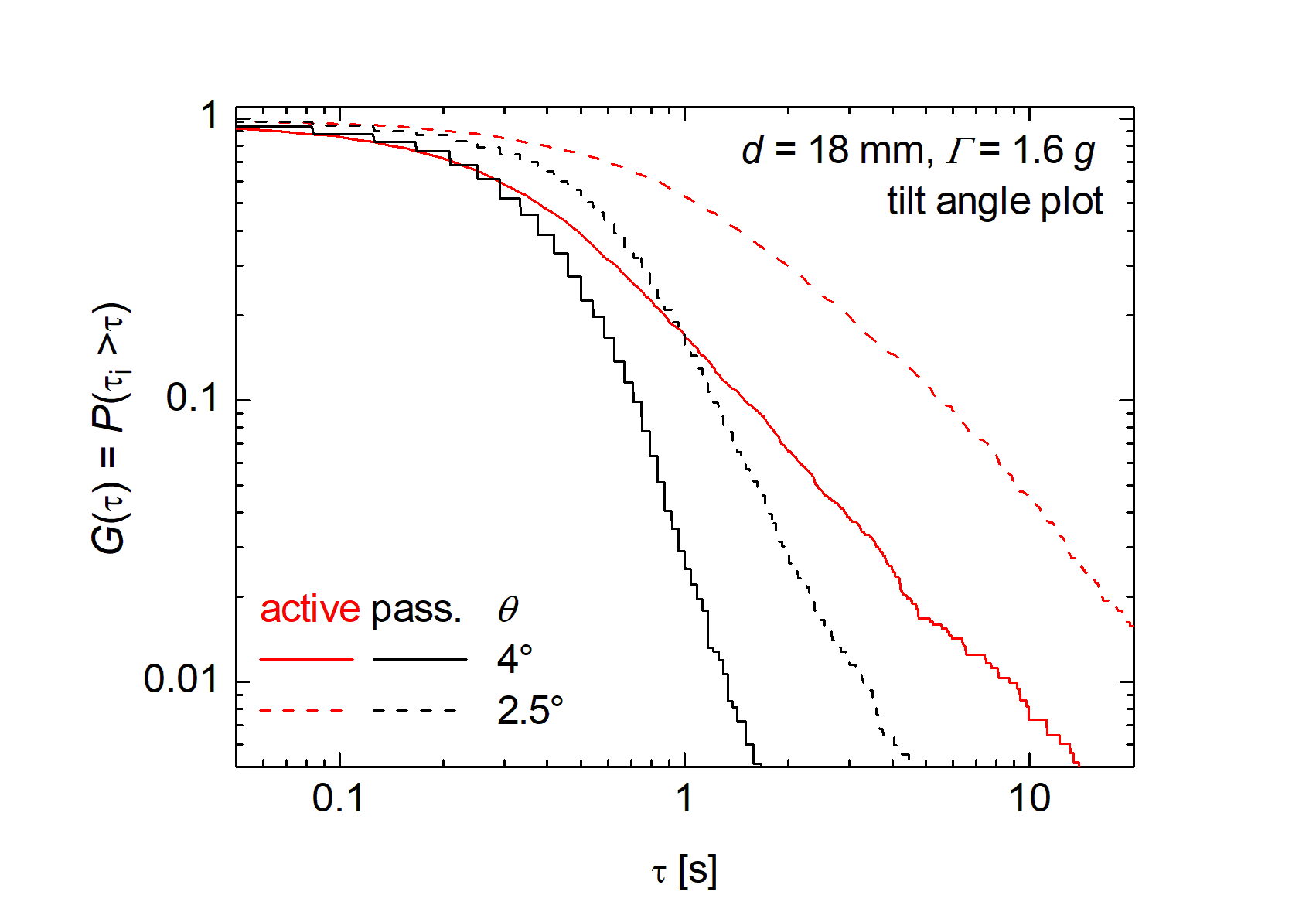}
      \caption{Tilt angle dependence of $G(\tau)$ for active and passive capsules. The main effect of tilt angle changes is a reduction of $\tau_0$.
      \label{fig:tilt}}
\end{figure}

The passage experiments were performed for three different opening sizes, three different excitation strengths and several tilt angles. In order to restrict the data in this paper to the essentials, we present as examples an excitation strength dependence at fixed tilt and orifice (Fig.~\ref{fig:exc}),
an orifice size dependence with fixed tilt and excitation (Fig.~\ref{fig:ori}),  and a tilt angle dependence at fixed orifice and excitation parameters (Fig.~\ref{fig:tilt}).

The influence of different excitation strengths on the
distribution of delay times between passages of individual particles is shown in Fig.~\ref{fig:exc}a for active and passive capsules.
It is seen that the influence of a higher or lower excitation strength on the global shapes of the curves is limited to a shift to longer or shorter times, respectively. With larger excitation amplitudes, the curves are shifted to shorter delay times $\tau_0$ but the slopes remain the same within experimental uncertainty. This means in particular that the exponents $\gamma$ of the fit curves, reflected in the slopes in the log-log plot for large $\tau$, are practically independent of the excitation strength. These $\gamma$ are shown for different orifice sizes and excitation strengths in
Fig.~\ref{fig:ori}b.
The fit parameters $\tau_0$ are systematically smaller for the passive particles than for the active ones at given $d$ and $\Gamma$ and $\theta$ for all experiments.
It is also seen that the exponents for passive and active capsules are substantially different. The average coefficient for active grains is $\gamma_{\rm a} =1.57 \pm 0.1$, whereas the average for passive grains is $\gamma_{\rm p} =2.58 \pm 0.05$. Both coefficients are large enough to guarantee the convergence of the integral in Eq.~(\ref{eq:tau_mean}).
The corresponding mean delay times $\bar\tau$ are $\approx 0.58~\tau_0$ for our active and $\approx 0.36~\tau_0$ for our passive particles.

Figure~\ref{fig:tilt} shows the tilt angle dependence of the delay distributions $G(\tau)$. The orifice width and excitation strength were kept constant. It is obvious that the slopes at large $\tau$ are only slightly influenced by the
variation of the effective gravitation ('motivation'). A change of the tilt angle $\theta$ from $2.5^\circ$ to $4^\circ$ reduces $\gamma$ by only 10\% for active and 7\% for passive capsules. On the other hand, the larger tilt angle reduces $\tau_0$ by 65\% for the active grains and 33\% for the passive ones.

\section{Discussion and summary}

The investigation of symmetrically and asymmetrically loaded spherocylindrical capsules on a vibrating plate can be used to explore the characteristics of self-propelled objects with qualitatively different features.
We have found that capsules equipped with an asymmetric weight perform a directed motion along their long axis with velocities between 5 mm/s and 10 mm/s. The directed motion is in the direction of the heavier part.  Velocities of individual capsules vary within a narrow range. For excitation strengths between 1.4 $g$ and 1.8 $g$ acceleration amplitude (70 Hz frequency), an increase of the excitation strength causes a faster motion. Above 1.8 $g$, the particles start to lift their lighter halves from the vibrating plate, and the propagation velocity drops.
At the same time, the capsules undergo fluctuations of their orientations on the scale of a few seconds, so that the persistence length of their
individual trajectories is of the order of a few particle lengths. This behavior reflects some of the features of bacteria, worms or ants in the search for food. When we break the in-plane symmetry of the vibrating plate by a slight tilt of $2^\circ$ to about $5^\circ$, the active motion of the particles is superimposed by a drift down the slope. There is a small tendency of individual active capsules to orient with the heavier part uphill so that the active motion counteracts the downward drift. When the active particles are enclosed in the container at large packing fractions, and they have contact to their neighbors, this tendency is less pronounced (see video in  \cite{SI2}). Thus, we assume that this effect is not the primary reason for the observed unique discharge characteristics of the container.

The symmetrically loaded, passive particles perform no directed motion, they show a purely diffusive behaviour, where the diffusion along the capsule axis is dominating and diffusion perpendicular to it can be practically neglected. They reflect the typical behavior of inanimate matter.
The comparison of both types of objects
may help to identify some statistical consequences of biological and granular particle activity.

Besides the motion of individual particles, among the interesting problems is the interaction of multiple particles, alignment and local order as a function of the packing density, the sedimentation of active particles, and the passage around obstacles or through bottlenecks. Our study was focused on the latter problem.
The first surprising observation is the discharge rate vs. orifice width. Passive capsules behave just as intuitively predicted, with the discharge rate roughly proportional to the orifice width. This means that the capsules leave the orifice at some average velocity that does not depend on the orifice width. Such a behaviour is comparable to many egress situations with living objects. In contrast, the outflow rate of the active capsules follows an orifice width dependence with at least square power. The exponent determined from the experiments is, independent of excitation strength and tilt
angle, $\beta=2.2$ within 10\% experimental uncertainty.
The reason for this is not obvious and may require further experimental studies and, primarily, appropriate numerical simulations.

The second difference between the discharge characteristics of our active and passive particles is the distribution of delay times $\tau$ between subsequent passages of particles. The cumulative functions $G(\tau)$ of passive capsules are characterized by an exponent $\gamma_{\rm p} = 2.58\pm 0.1 $ of the long-time tail, whereas the active capsules exhibit a much weaker exponent of  $\gamma_{\rm a} = 1.57\pm 0.15$. This means that their probability of long delays is considerably higher than for the passive particles. Compared to animals passing a gate \cite{Zuriguel2014} or pedestrians exiting through a door \cite{Garcimartin2016}, the exponents $\gamma$ in our experiments are considerably smaller, i. e. the probability of long-term congestions is much larger for our active capsules.
One of the reasons is that the activity is not directed towards the outlet like in usual egress experiments
with living matter, but the individual particle velocities are determined by the 'erratic' orientations of the individuals. This behavior can at best be compared to random motion of lower organisms or
a crowd in disordered panic.

In vibrated hoppers, Janda et al. \cite{Janda2009b} determined time lapses between subsequent grain passages and found tails with exponents near $\tau^{-2}$ in the $p(\tau)$ graphs (corresponding to asymptotic  tails $\propto \tau^{-1}$ in the cumulative distributions). In some of these experiments, the integral to calculate the mean time lapse diverges and one needs an alternative characterization of the dynamic state of the system.
An experiment with 'bristlebots', self-propelled elongated mechanical toys \cite{Patterson2017,Patterson_comment}, yielded exponents that are also much smaller than those found in our study, close to the critical value of $\gamma_{\rm c} =1$ where $\bar \tau$ diverges.

Finally,
suspensions of particles passing bottlenecks show a similar intermittent behavior, but with tails consistently flatter than $\tau^{-2}$ in $p(\tau)$~\cite{Souzy2020} ($\gamma <1$).
With respect to the intermittent character of the flow,
the present system thus represents an
intermediate dynamics between
those of living objects and those of
suspensions in bottleneck geometries.

The results obtained for all these experimental parameters are surprisingly robust. Changes of the tilt angle (the 'incentive') and the excitation parameters do essentially affect the time scales of the egress, but leave the scaling exponents $\gamma$ and $\beta$ practically unaffected. This suggests that these dynamical parameters are determined essentially by the particle dynamics, not by the geometrical or dynamical  details. The opening size merely determines the outflow rate but not the shapes of the characteristic curves. This makes us confident that many of the above results can be generalized.


\section*{Acknowledgements}
We cordially thank C. Wagner for leaving the shaker device to us.
The authors acknowledge DLR for funding within project  EVA (50WM2048). M. M. acknowledges a research grant from the Iranian
Ministry of Science, Research and Technology for funding part of her stay in Magdeburg.

\section*{Appendix A: Distributions of displacements for active and passive particles}

\begin{figure}[htbp]
\centering
     \includegraphics[width=0.75\columnwidth]{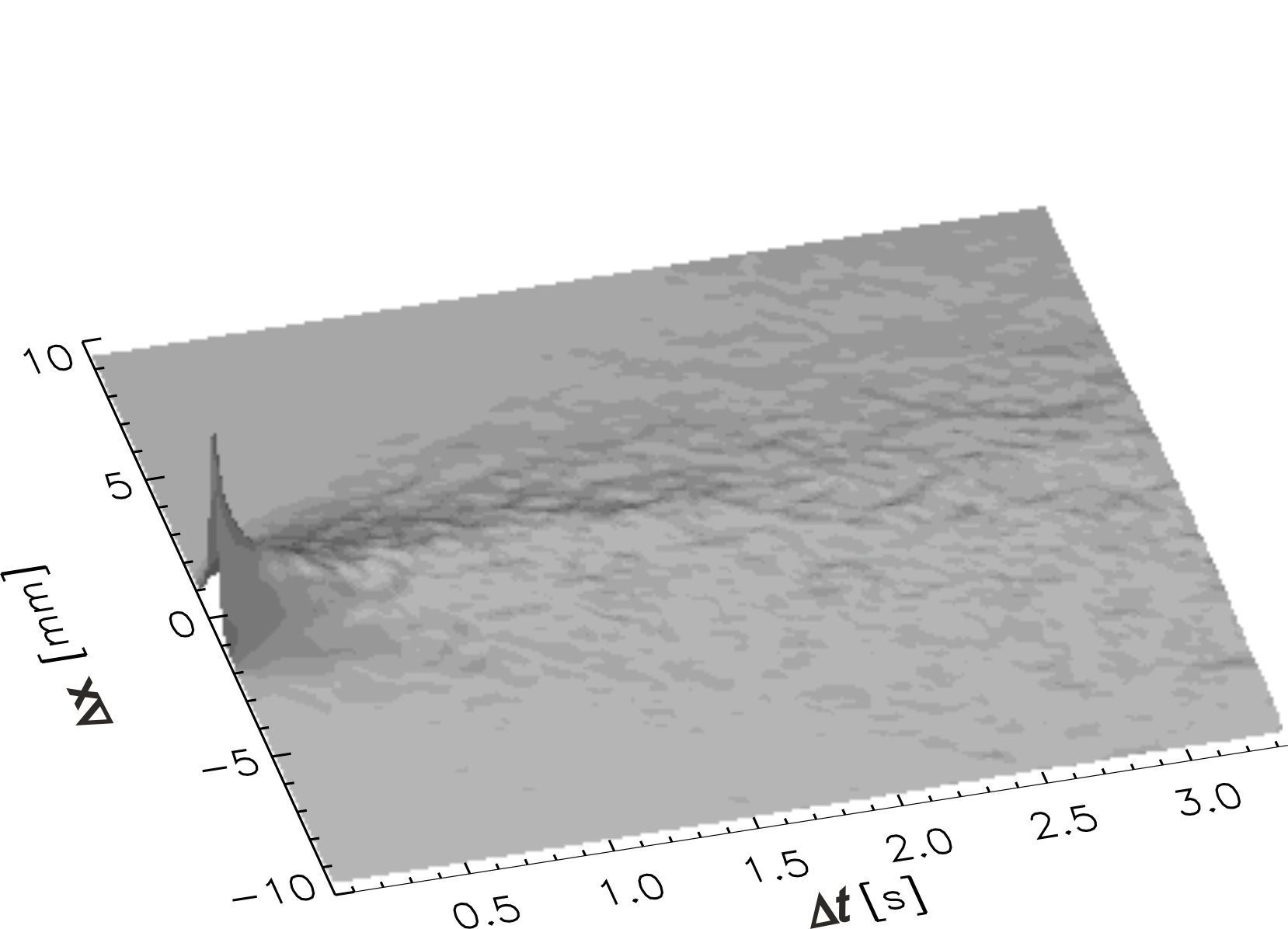} \\
     \includegraphics[width=0.75\columnwidth]{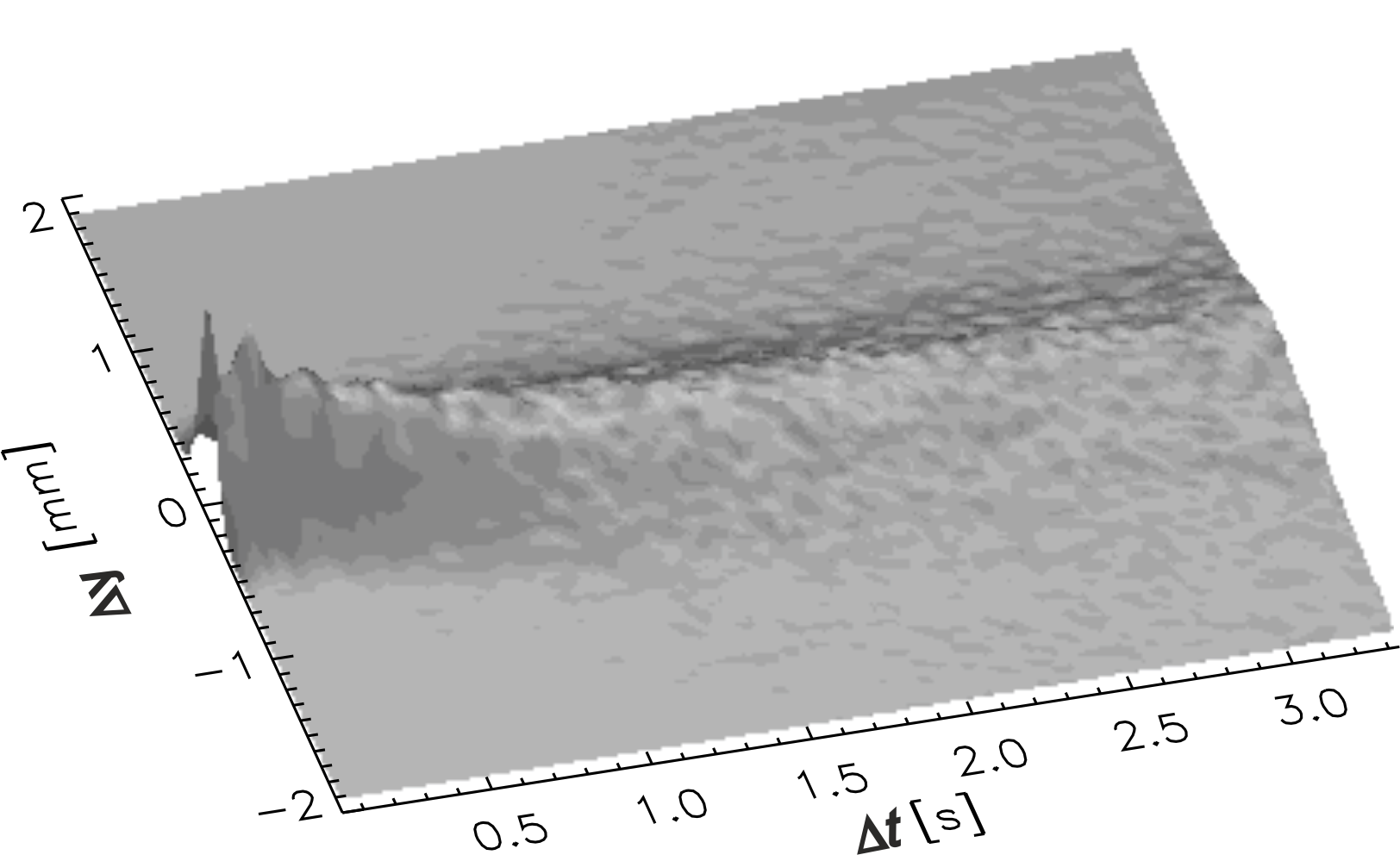} \\
     \includegraphics[width=0.75\columnwidth]{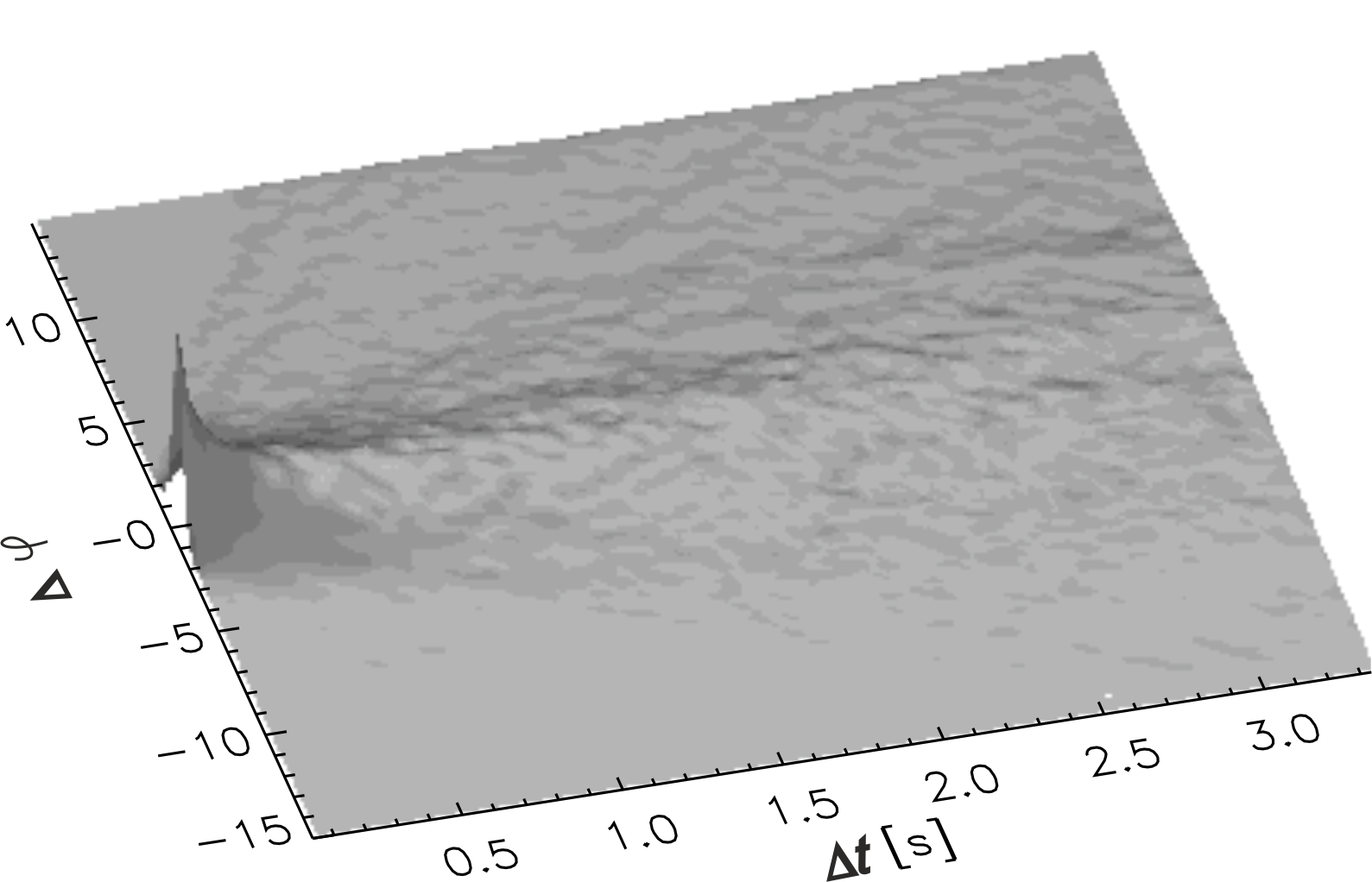}\\
      \caption{Top to bottom: Distributions of displacements of a passive capsule along the capsule axis, of the displacement perpendicular to the capsule axis, and of the rotation angles as a function of the delay $\Delta t$. $d=18$~mm, $\theta=4^\circ$ \label{fig:A1}}
\end{figure}

Figure \ref{fig:A1} shows the distribution of translational and rotational displacements of the passive particles, measured from 11 individual trajectories.
It is obvious that the distribution functions broaden continuously, leading to the mean values shown in Fig. \ref{fig:msd} above, but the spatial displacements are statistically symmetric with respect to positive and negative steps. Note that the diffusion along the particle axis is much faster than in perpendicular direction, by a factor of approximately five. The rotational distribution also broadens continuously, but it has in general a small bias towards one of the two rotation senses. This small asymmetry is not considered relevant here, each individual capsule has
a slight bias in one or the other direction.

\begin{figure}[htbp]
\centering
      \includegraphics[width=0.75\columnwidth]{active_x14n.png}
      \includegraphics[width=0.75\columnwidth]{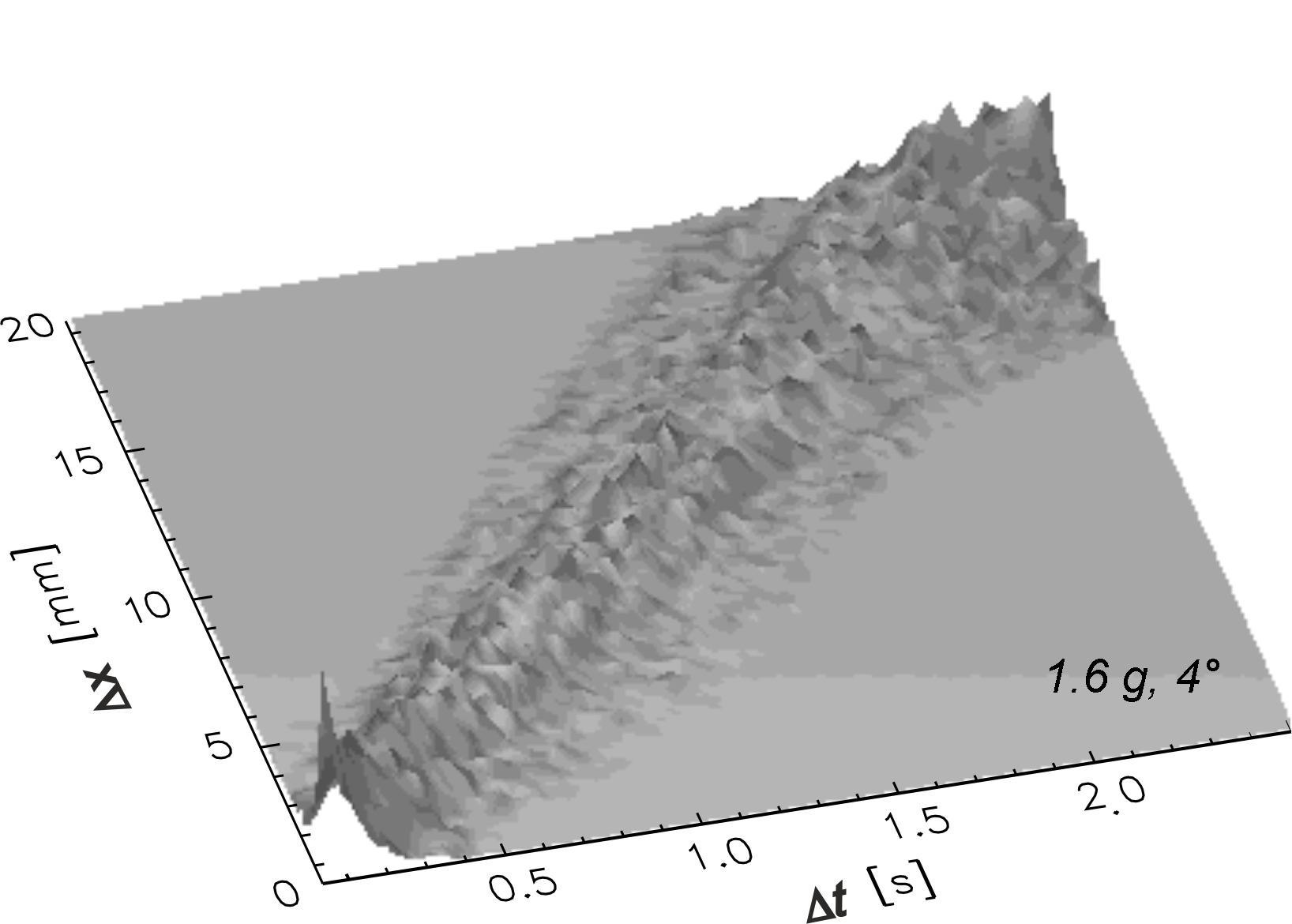}
      \includegraphics[width=0.75\columnwidth]{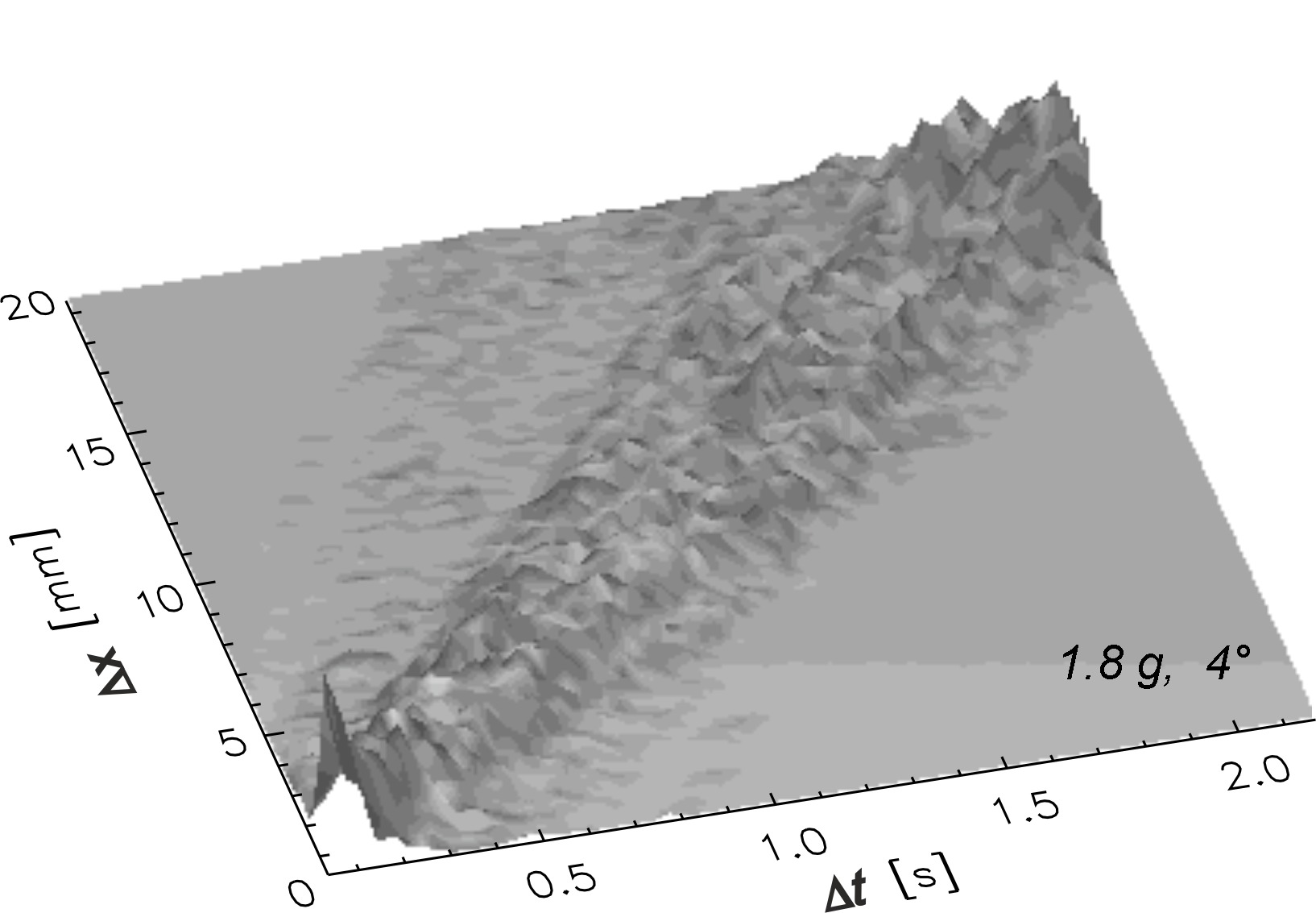}
      \caption{Distributions of displacements of an active capsule
      along its axis as a function of the delay $\Delta t$, for an excitation with accelerations
      1.4 $g$, 1.6 $g$ and 1.8 $g$ from top to bottom.
       (The upper graph is equivalent to Fig.~\ref{fig:hist}, left).
      \label{fig:A2}}
\end{figure}

Figure \ref{fig:A2} shows the distribution of translational displacements of the active particles along their axis, measured from 11 individual trajectories each.
The distribution functions broaden continuously with larger delay $\Delta t$, but most prominent is the shift of the funcion in the direction of the front end of the particle.
This shift becomes faster with stronger excitation.
The displacement perpendicular to the active axis are comparable to those of the
passive capsules.

\pagebreak
\section*{Appendix B: Fit of the cumulative distribution $G(\tau)$ }

\begin{figure}[htbp]
\centering
      \includegraphics[width=0.75\columnwidth]{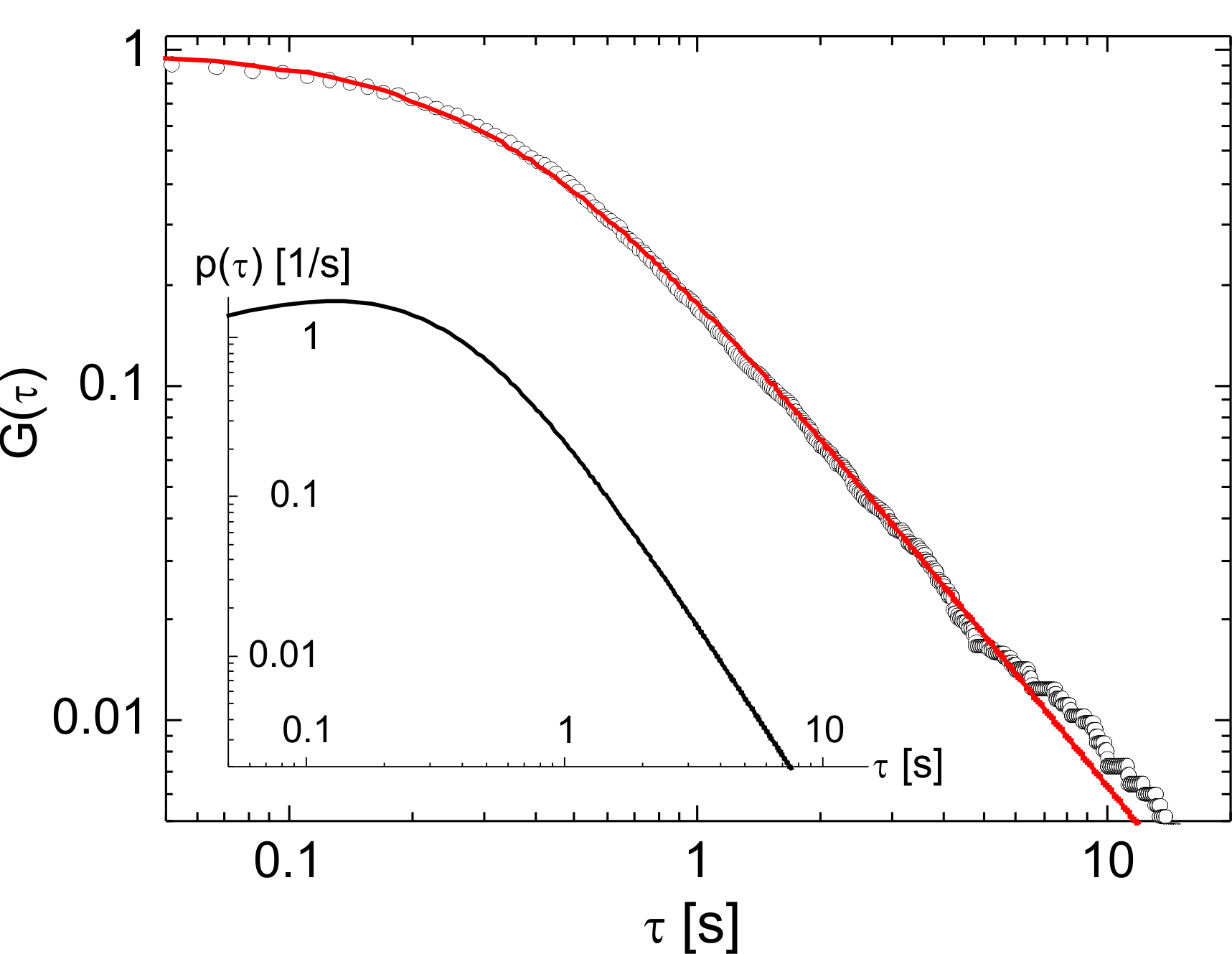}
       \caption{Cumulative distribution of the delays between passages of subsequent capsules for $d=18$~mm, $\theta=4^\circ$ and $\Gamma=1.6 ~g$ and the fit to Eq.~(\ref{eq:tau2}) (red curve) with parameters $\gamma=1.52$ and $\tau_0=0.36$~s. The inset shows the distribution $p(\tau)$ that follows from Eq.~(\ref{eq:tau3}) using the above fit parameters.
      \label{fig:B1}}
\end{figure}

Figure \ref{fig:B1} demonstrates the quality of the fit function, Eq.~(\ref{eq:tau2}), for the
determination of characteristic parameters describing the experiment.
The experimental data are well reproduced by the empirical fit.
There is a slight deviation for very long $\tau$ which arises primarily from a few particles dwelling in the container when it is nearly emptied. One can get an impression of the distribution function $p(\tau)$ using the fit parameters and Eq.~(\ref{eq:tau3}). It agrees with the experimental data satisfactorily, but naturally the experimental results for $p(\tau)$ scatter much more than those of its integral $G(\tau)$.

\end{document}